\documentclass[a4paper,11pt]{article}
\pdfoutput=1 

\usepackage{jcappub} 

\usepackage[T1]{fontenc} 
\usepackage[bottom]{footmisc}
\usepackage{subcaption}
\usepackage{changes} 

\title{\boldmath Chameleon Screening Depends on the Shape and Structure of NFW Halos}

\author[a]{Andrius Tamosiunas,}
\author[a]{Chad Briddon,}
\author[a]{Clare Burrage,}
\author[b]{Weiguang Cui,}
\author[a]{and Adam Moss}

\affiliation[a]{School of Physics and Astronomy, University of Nottingham, Nottingham, NG7 2RD,\\ United Kingdom}

\affiliation[b]{Institute for Astronomy, University of Edinburgh, Edinburgh, EH9 3HJ, United Kingdom }

\emailAdd{andrius.tamosiunas@nottingham.ac.uk}
\emailAdd{chad.briddon@nottingham.ac.uk}
\emailAdd{clare.burrage@nottingham.ac.uk}
\emailAdd{Weiguang.Cui@ed.ac.uk}
\emailAdd{adam.moss@nottingham.ac.uk}

\abstract{Chameleon gravity is an example of a model that gives rise to interesting phenomenology on cosmological scales while simultaneously possessing a screening mechanism, allowing it to avoid solar system constraints. Such models result in non-linear field equations, which can be solved analytically only in simple highly symmetric systems. In this work we study the equation of motion of a scalar-tensor theory with chameleon screening using the finite element method. More specifically, we solve the field equation for spherical and triaxial NFW cluster-sized halos. This allows a detailed investigation of the relationship between the NFW concentration and the virial mass parameters and the magnitude of the chameleon acceleration, as measured at the virial radius. In addition, we investigate the effects on the chameleon acceleration due to halo triaxiality. We focus on the parameter space regions that are still allowed by the observational constraints. We find that given our dataset, the largest allowed value for the chameleon-to-NFW acceleration ratio at the virial radius is $\sim 10^{-7}$. This result strongly indicates that the chameleon models that are still allowed by the observational constraints would not lead to any measurable effects on galaxy cluster scales. Nonetheless, we also find that there is a direct relationship between the NFW potential and the chameleon-to-NFW acceleration ratio at the virial radius. Similarly, there is a direct (yet a much more complicated) relationship between the NFW concentration, the virial mass and the acceleration ratios at the virial radius. Finally, we find that triaxiality introduces extra directional effects on the acceleration measurements. These effects in combination could potentially be used in future observational searches for fifth forces.}

\begin{document}
\maketitle
\flushbottom

\section{Introduction}
\label{introduction}
\raggedbottom

Scalar-tensor models of gravity offer a novel approach to tackling some of the key issues in modern cosmology \cite{Faraoni2004}. Such models introduce an extra scalar degree of freedom, which often results in an interesting phenomenology on cosmological scales \cite{Fujii2003,Benisty2021}. However, a major issue of such models is fitting the stringent solar system observational constraints. In this context, the subset of scalar-tensor theories possessing a screening mechanism is of special interest. Models possessing chameleon, symmetron or Vainshtein screening offer the possibility of fitting the tight local constraints, while still possessing interesting phenomenology on different scales \cite{Vainshtein1972, Khoury2004, Hinterbichler2011b}.
This work will focus on the chameleon model, which introduces a non-minimally coupled scalar field that results in a fifth force dependent on the ambient mass density (see Ref.~\cite{Burrage2018} for a review of the theory and the current observational constraints).  

Scalar fields coupled to gravity appear in many extensions and UV completions of general relativity (GR), e.g. various string theory models \cite{Hinterbichler2011}. The subset of such models that are able to self-screen are then of particular interest in the context of modern problems in cosmology. However, single scalar field models possessing a screening mechanism, such as the chameleon and the symmetron, are well constrained \cite{Jain2013, Lombriser2014, Koyama2016, Sakstein2016, Burrage2016, nunes2017, Pernot-Borr2021}. What is more, it has been shown that such models cannot simultaneously screen and self-accelerate cosmologically \cite{Wang2012}. Nonetheless, such models can still possess exciting phenomenology on a wide range of scales. Also, it has been recently shown that screening effects could also in principle be detected on laboratory scales \cite{ Burrage2014, elder2016, chiow2018, Sabulsky2019, tino2021}. Finally, understanding such relatively simple yet phenomenologically interesting models can be of great value when building more complex high-energy models as well as checking the validity of GR as the correct theory of gravity on different scales. 

Theories of modified gravity possessing screening lead to non-linear equations of motion that, in general, cannot be solved analytically, with an exception of a few highly symmetric cases. This becomes an issue in the astrophysical context, when studying the complex density distributions that characterize dark matter halos in galaxies and galaxy clusters. Different features of theories possessing chameleon screening have been studied extensively in previous work on different astrophysical scales \cite{Arnold2016, Wilcox2016_sim, Naik2018, Dima2021}. In this work we will specifically focus on galaxy clusters. Cluster-sized halos were chosen as they have historically played an important role in the dark matter-modified gravity debate \cite{Sanders2003, Clowe2006, Nusser2006}. Similarly, clusters, being the largest gravitationally-bound objects in the Universe, play an important role when studying structure formation. Likewise, the ability of measuring cluster masses in different ways (e.g. X-ray astronomy, weak lensing and the Sunyaev–Zeldovich effect) has been explored as a powerful tool for constraining various modified gravity models \cite{Lombriser2012,Terukina2013,Wilcox2015, Wilcox2016, Sakstein2016,Cataneo2018, Tamosiunas2019}. Clusters, having regions of both high densities (cluster cores) and low densities (cluster outskirts) are a perfect environment for testing chameleon gravity. As shown in previous work (e.g. Ref.~\cite{Wilcox2015}), even subtle chameleon gravity effects could in principle be detected by comparing X-ray and weak lensing profiles near the central core regions and in the outskirts of clusters. An important aspect of cluster physics in the context of screened modified gravity is that of cluster shapes. Here we address the effects of cluster shapes on chameleon screening by solving the chameleon equations for triaxial NFW halos.

One approach when solving the equations of motion resulting from the Navarro-Frenk-White (NFW) distributions is applying the finite element method (FEM) \cite{Turner1956,Navarro1997}. The FEM technique allows us to numerically solve the non-linear equations arising from complex, non-symmetric and time-varying density distributions. In this work we develop an FEM-based approach for numerically solving the chameleon field equation that describes the scalar field arising from static NFW cluster-sized halos. As a sample dataset we employ the simulated galaxy cluster catalog from the T{\scriptsize HE} T{\scriptsize HREE} H{\scriptsize UNDRED} P{\scriptsize ROJECT}\footnote{\url{https://the300-project.org}}, which contains 324 realistic galaxy clusters \cite{Cui2018}. In order to investigate the chameleon effects in NFW halos, we solve the field equation for a range of virial masses and concentration parameters. 

\begin{figure}
  \centering
    \includegraphics[width=0.60\textwidth]{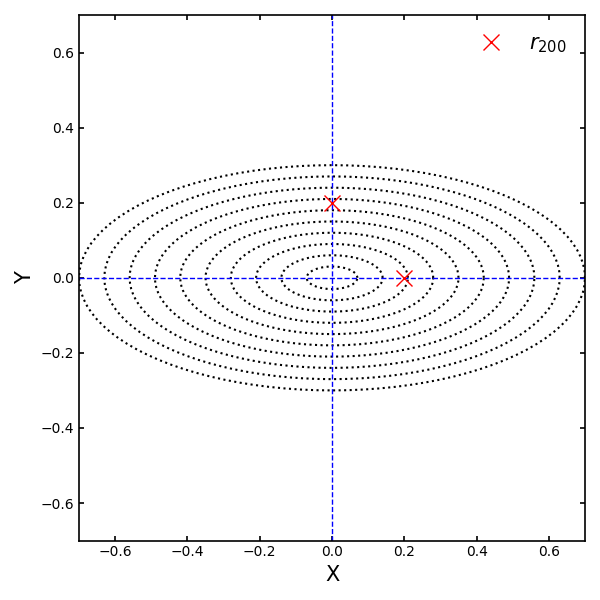}
    \caption{Effects of triaxiality on equidistant measurements along the X and the Y axes. The dotted lines correspond to the contours of equal density. The red crosses correspond to an equal distance along the X and the Y axes (here set to the virial radius $r_{200}$ of the halo). The contours were produced using axis ratios: $q_{a} = 0.3$ and $q_{b} = 0.7$. The triaxiality effects result in the chameleon acceleration at the virial radius having significantly different values along the X and the Y axes.} 
    \label{triaxiality_effects}
\end{figure}

In this work we also investigate NFW halo triaxiality, which introduces directional effects that could potentially be employed in the observational searches for the chameleon and similar models. More specifically, we model the cluster halos using a triaxial NFW distribution. The triaxial NFW profile, when compared to the spherical NFW distribution in 2D, looks elongated along one axis, while being compressed along the other. Hence, if we measure the chameleon acceleration at the same distance from the centre of the halo along the X and the Y axes, we expect to obtain significantly different acceleration values, as we are gauging different parts of the triaxial NFW density profile with different gradients. We expect the acceleration difference $\Delta a_{\phi}$ to be proportional to the difference between the axis ratios $|q_{a} - q_{b}|$ (see eq.~(\ref{triaxial_radius})). We also expect the acceleration difference to vary depending on the angle at which it is measured, with the measurements along the X and the Y axes resulting in maximum difference. Figure \ref{triaxiality_effects} illustrates this effect. In order to study these triaxiality effects, we calculate the chameleon-to-NFW acceleration ratio in triaxial coordinates and study how the results depend on the axis ratios $q_{a}$ and $q_{b}$. To investigate the triaxiality effects described in figure \ref{triaxiality_effects}, we obtain the acceleration ratio difference along the X and the Y axes.

In the following sections we introduce the FEM technique used, summarize the key results and outline the implications for observational chameleon fifth force searches. More specifically, section \ref{chameleon_introduction} introduces the theory behind the chameleon mechanism. In section \ref{nfw_halos} we discuss the properties of the NFW density distributions. Section \ref{The_dataset} describes the used simulated galaxy cluster dataset in detail. Section \ref{FEM_introduction} summarizes the key features of the FEM approach. Section \ref{spherical_results} summarizes the results for the spherical NFW halos. Section \ref{triaxiality_results} outlines the effects of triaxiality. And finally, appendix \ref{appendix_analytical_approx} summarizes the approximate analytical predictions. 

In this work we use natural units ($c = \hbar = 1$) and a positive metric signature, i.e. $\eta_{\mu \nu}=\operatorname{diag}(-1,1,1,1)$. We denote the reduced Planck mass as $
M_{\rm Pl} = \left(8 \pi G\right)^{-1}$, with $G$ as Newton's gravitational constant. Quantities denoted by a tilde refer to the Jordan frame metric. Quantities denoted by the hat symbol are rescaled by the background value (see section \ref{FEM_introduction}). For calculations including the Hubble parameter we use the following value: $h = 0.6777$. Finally, all the figures (unless otherwise specified) were produced using the following NFW halo inner cutoff radius $r_{\rm cut} = 0.15 \times r_{200}$.

\section{The Chameleon Model}
\label{chameleon_introduction}

The chameleon model refers to a non-minimally coupled scalar-tensor theory described by the following action:

\begin{equation}
S=\int \mathrm{d} x^{4} \sqrt{-g}\left(\frac{M_{\rm Pl}^{2}}{2} R-\frac{1}{2} \nabla_{\mu} \phi \nabla^{\mu} \phi-V(\phi)\right)-\int \mathrm{d} x^{4} \mathcal{L}_{m}\left(\varphi_{m}^{(i)}, \tilde{g}_{\mu \nu}^{(i)}\right),
\end{equation}

\noindent with $\phi$ as the scalar field, $V(\phi)$ as the potential, $\mathcal{L}_{m}$ as the matter Lagrangian, $\varphi_{m}^{(i)}$ as the matter fields and $\tilde{g}_{\mu \nu}^{(i)}$ as the Jordan frame metric. The quantities denoted by the superscript \textit{i} refer to the i-th matter species. The Jordan frame metric can then be related to the Einstein frame metric via the rescaling factor $A_{i}$:

\begin{equation}
\tilde{g}_{\mu \nu}^{(i)}=A_{i}^{2}(\phi) g_{\mu \nu}.
\end{equation}

\noindent The coupling is then given by: 

\begin{equation}
\beta_{i}=M_{\rm Pl} \frac{\mathrm{d}\left(\ln \left(A_{i}\right)\right)}{\mathrm{d} \phi}.
\end{equation}

\noindent Varying the action with respect to the scalar field gives the equation of motion:

\begin{equation}
\square \phi=V_{, \phi}-\sum_{i} \frac{1}{M_{i}} T_{\mu \nu}^{(i)} \tilde{g}_{(i)}^{\mu \nu},
\label{EOM}
\end{equation}

\noindent with $T_{\mu \nu}^{(i)}$ as the energy-momentum tensor and $M_{i} \equiv M_{\rm Pl}/\beta_{i}$. Using the right hand side of eq.~(\ref{EOM}) one can define the effective potential:

\begin{equation}
V_{\rm eff}(\phi)=V(\phi)+\sum_{i} \frac{ \rho_{i}}{M_{i}} \phi,
\label{effective_potential}
\end{equation}

\noindent where we have taken into account that for non-relativistic matter $T_{\mu \nu}^{(i)} \tilde{g}_{(i)}^{\mu \nu} \approx-\rho_{i}$. The shape of the density-dependent effective potential controls the effective mass of the chameleon field in such a way that the fifth force effects are suppressed in high-density environments. More specifically, the effective mass is given by: 

\begin{equation}
m_{\phi}^{2}=V_{\rm eff, \phi \phi}\left(\phi_{\rm min}\right),
\end{equation}

\noindent where $\phi_{\rm min}$ is the minimal field value, obtained by setting the first derivative of the effective potential to 0. The effective mass is related to the Compton wavelength via: $\lambda_{C} = m_{\phi}^{-1}$. 

A common choice for the bare potential $V(\phi)$ is the following:

\begin{equation}
V(\phi)=\Lambda^{4}\left(1+\frac{\Lambda^{n}}{\phi^{n}}\right),
\label{chameleon_potential}
\end{equation}

\noindent with $n$ as an integer and $\Lambda$ as the energy scale (often set to the dark energy scale $\Lambda = 2.4$ meV). Given our choice of the bare potential, the background value of the field (for the case of a single matter species) is given by:

\begin{equation}
\phi_{\rm \infty}=\left(\frac{n M \Lambda^{n+4}}{ \rho_{\rm \infty}}\right)^{\frac{1}{n+1}},
\label{phi_min}
\end{equation}

\noindent where $M \equiv M_{\rm Pl}/\beta$ and $\rho_{\rm \infty}$ is the background density (i.e. the mean density of the Universe in our case).

Assuming a single matter species and a time-independent scalar field and matter distribution, the field equation is finally given by: 

\begin{equation}
\nabla^{2} \phi=-\frac{n \Lambda^{n+4}}{\phi^{n+1}}+\frac{ \rho}{M}.
\label{field_equation}
\end{equation}

\noindent By solving eq.~(\ref{field_equation}) for some density distribution $\rho$, one can deduce the fifth force due to the chameleon field on a test particle with mass $m$: 

\begin{equation}
F_{\phi}=-\frac{m}{M } \nabla \phi.
\label{fifth_force}
\end{equation}

\section{The NFW Halo Properties}
\label{nfw_halos}

The dark matter halos on galaxy and galaxy cluster scales are known to follow a universal density profile. A common choice for such a profile is the NFW profile \cite{Navarro1997, Wright1999, Wagner2020}:

\begin{equation}
\rho(r)=\frac{\rho_{0}}{\frac{r}{r_{s}}\left(1+\frac{r}{r_{s}}\right)^{2}}.
\label{NFW_profile}
\end{equation}

\noindent Here $r_{s}$ is the characteristic scale radius, such that $r_{200} = c_{\rm NFW} r_{s}$. The virial radius $r_{200}$ refers to the radius at which the density is equal to 200 times the critical density and $c_{\rm NFW}$ is the concentration parameter. The term in the numerator refers to $\rho_{0} = \delta_{c} \rho_{c}$, with $\rho_{c}$ as the critical density and $\delta_{c}$ as the characteristic overdensity given by:

\begin{equation}
\delta_{c}=\frac{200}{3} \frac{c_{\rm NFW}^{3}}{\ln (1+c_{\rm NFW})-c_{\rm NFW} /(1+c_{\rm NFW})}.
\end{equation}

\noindent Then the mass enclosed between two radii $r_{\rm A}$ and $r_{\rm B}$ is  given by:

\begin{equation}
\begin{aligned}
M_{\mathrm{NFW}}\left(r_{\rm A}, r_{\rm B}\right)=& \int_{r_{\rm A}}^{r_{\rm B}} 4 \pi r^{2} \rho(r) \mathrm{d} r=\\
& 4 \pi \rho_{0} r_{s}^{3}\left[\frac{r_{s}\left(r_{\rm A}-r_{\rm B}\right)}{\left(r_{s}+r_{\rm B}\right)\left(r_{s}+r_{\rm A}\right)}+\ln \left(\frac{r_{s}+r_{\rm B}}{r_{s}+r_{\rm A}}\right)\right].
\end{aligned}
\end{equation}

\noindent The virial mass (i.e. the mass enclosed by $r_{200}$) is then defined as :

\begin{equation}
M_{200}=\frac{800 \pi}{3} r_{\mathrm{200}}^{3} \rho_{\mathrm{c}}.
\label{virial_mass_spherical}
\end{equation}

\noindent Assuming the NFW profile, the regular Newtonian potential is given by:

\begin{equation}
\Phi_{\mathrm{NFW}}(r)=-\frac{4 \pi G \rho_{0} r_{s}^{3}}{r} \ln \left(1+\frac{r}{r_{s}}\right).
\label{NFW_pot}
\end{equation}

\noindent In this work we introduce an inner density cutoff, which sets the density to be constant inside a radius $r_{\rm cut}$ (see section \ref{FEM_introduction} for a wider discussion). This is done in order to avoid numerical problems at the central point ($r \rightarrow 0$), where the density grows indefinitely. In addition, the density profiles of galaxy clusters are observed to flatten out in the central region (so that the halo is said to have a core) \cite{Sarazin1986}. The NFW mass with an inner cutoff is given by:

\begin{equation}
\begin{aligned}
M_{\mathrm{NFW}}^{\mathrm{cut}}(<r)=& \int_{0}^{r_{\text {cut }}} 4 \pi r'^{2} \rho(r') \mathrm{d}r'+\int_{r_{\text {cut }}}^{r} 4 \pi r'^{2} \rho(r') \mathrm{d}r' =&\\ 
\frac{4 \pi \rho_{0}}{r_{\mathrm{cut}} / r_{s}\left(1+r_{\mathrm{cut}} / r_{s}\right)^{2}} \frac{r_{\mathrm{cut}}^{3}}{3} \; +& \; 4 \pi \rho_{0} r_{s}^{3}\left[\frac{r_{s}\left(r_{\mathrm{cut}}-r\right)}{\left(r_{s}+r\right)\left(r_{s}+r_{\mathrm{cut}}\right)}+\ln \left(\frac{r_{s}+r}{r_{s}+r_{\mathrm{cut}}}\right)\right],
\end{aligned}
\label{M_cut}
\end{equation}

\noindent with $r_{\rm cut}$ as the cutoff radius. The potential with a cutoff is then given by:

\begin{equation}
\Phi_{\mathrm{NFW}}^{\rm cut}(r)=-\frac{4 G \pi \rho_{0} r_{s}^{3}}{3 r}\left[\frac{r_{\mathrm{cut}}\left(4 r_{\mathrm{cut}}+3 r_{s}\right)}{\left(r_{\mathrm{cut}}+r_{s}\right)^{2}}+3 \ln \left(\frac{r+r_{s}}{r_{\mathrm{cut}}+r_{s}}\right)\right].
\end{equation}

\noindent Finally, the gravitational acceleration sourced by the matter in an NFW halo is given by:

\begin{equation}
    a_{\rm NFW}(r) = - \frac{\mathrm{d} \Phi_{\rm NFW}^{\rm cut}(r)}{\mathrm{d}r} = -\frac{G M_{\rm NFW}^{\rm cut}(<r)}{r^{2}}.
\end{equation}

The equations above describe spherical NFW halos. However, in nature most halos have a complex density distribution. Generally, this complex distribution is described more accurately by a triaxial NFW profile than by a spherical one \cite{Jing2002, Limousin2013, Tchernin2020}. Specifically, triaxial halos can be described by switching to triaxial coordinates: 

\begin{equation}
    R^{2} = \frac{x^{2}}{q_{a}^{2}} + \frac{y^{2}}{q_{b}^{2}} + \frac{z^{2}}{q_{c}^{2}},
    \label{triaxial_radius}
\end{equation}

\noindent with $x$, $y$, $z$ as the Cartesian coordinates and $q_{a}$, $q_{b}$, $q_{c}$ as the axis ratios defined such that $q_{a} < q_{b} < q_{c} = 1$. Under this convention, the triaxial density can be found by switching from the radial to the triaxial coordinates in eq.~(\ref{NFW_profile}): $\rho(r) \rightarrow \rho(R)$. Similarly, the triaxial mass is then given by:

\begin{equation}
M_{\rm NFW}^{\rm ta}(<R)=4 \pi q_{a} q_{b} \int_{0}^{R} \rho(R') R'^{2} \mathrm{d}R'.
\end{equation}

\noindent This results in the triaxial virial mass being given by:

\begin{equation}
M_{200} = (800\pi/3)q_{a}q_{b}R_{200}^{3}\rho_{c}, 
\label{triaxial_M200}    
\end{equation}

\noindent with the virial triaxial radius defined such that $R_{200} = (q_{a}q_{b})^{-1/3} r_{200}$. Analogously the triaxial mass with a cutoff is calculated by switching to triaxial coordinates and multiplying eq.~(\ref{M_cut}) by the axis ratios.

\section{The Dataset}
\label{The_dataset}
\raggedbottom

In order to solve the chameleon field equations for a realistic set of NFW halos, we used T{\scriptsize HE} T{\scriptsize HREE} H{\scriptsize UNDRED} P{\scriptsize ROJECT} dataset \cite{Cui2018}. T{\scriptsize HE} T{\scriptsize HREE} H{\scriptsize UNDRED} P{\scriptsize ROJECT} refers to a simulation suite of 324 galaxy clusters taken from a dark matter only MDLP2 Multi Dark simulation with the following \textit{Planck} cosmological parameters: $\Omega_{M} = 0.307$, $\Omega_{B} = 0.048$, $\Omega_{\Lambda} = 0.693$, $h = 0.6777$, $\sigma_{8} = 0.823$ and $n_{s} = 0.96$ \cite{Klypin2016}. The individual cluster data was obtained by taking the 324 most massive clusters (at $z = 0$) and tracing the particles in the region of 15 $h^{-1}$ Mpc around the central point to their original position. The clusters were then re-simulated to include full baryonic effects by splitting the dark matter particles into dark matter and gas particles with
masses set by the baryonic matter fraction of the Universe (see Ref.~\cite{Cui2018} for a detailed description of the re-simulation procedure). The re-simulation was carried out using the {\scriptsize GADGET}X code \cite{Rasia2015}. The 324 cluster dataset has been recently used to investigate the density profile evolution with redshift, the substructure of galaxy clusters and of surrounding galaxy groups as well as the properties of backsplash galaxies in simulations of clusters \cite{Mostoghiu2019, Li2020, Haggar2020, Haggar2021}. 

The halo properties for each of the 324 clusters were deduced using Amiga's Halo Finder (AHF) \cite{Gill2004, Knollmann2009}. AHF uses the simulation outputs to locate the overdensities in an adaptively smoothed density field in order to locate halos and their corresponding substructures. For each located halo AHF calculates the key quantities such as the virial mass and radius, the peculiar velocity, the position of the rotation curve maximum, the maximum velocity of the rotation curve and other useful quantities. In addition, the AHF calculates a number of radial profiles such as the density and the overdensity along with the rotation curves. Finally, AHF infers the shape of the cluster in the form of the triaxial axes ratios. The key quantities for this work are the virial radius $r_{200}$ and the corresponding mass $M_{200}$ along with the concentration parameter $c_{\rm NFW}$, which are deduced by AHF for each cluster halo. More specifically, we used the catalogue of the 324 central halos for each of the re-simulated clusters.

A subtle, but important point is that the concentration parameter can be calculated in a number of different ways by the halo finder software. Specifically, the spherically averaged halo density distribution can be fitted directly by assuming the NFW profile. The concentration can then be deduced as the ratio between the virial radius and the characteristic radius: $c_{\rm NFW} = r_{200}/r_{s}$. One can also use another functional form of the density distribution, such as the Einasto profile. However, it has been found that using the Einasto profile results in fluctuations of the concentration parameter, which is due to smaller curvature of the profile. An alternative approach is to use a more profile-independent method, which deduces the concentration parameter from the ratio of the maximum circular velocity, $v_{\rm max}$, and the virial velocity $v_{\rm 200}$. The ratio $v_{\rm max}/v_{\rm 200}$ can be related to concentration without assuming the NFW profile (see section 3 in Ref.~\cite{Prada2012} for a detailed discussion on the different methods of calculating the concentration). As described in \cite{Prada2012}, the difference between the results of the two outlined methods for calculating the concentration parameter varies between $5-15 \%$. Note, however, that the difference of the density distribution deduced using the two methods mostly manifests in the central regions of the clusters. In our work we found that outside the central regions of the clusters, there is a good agreement between the two estimates for the concentration. Given that we introduce an inner density cutoff, we find a generally good agreement between the two ways of deducing the density profile. Furthermore, we compared the results for all the chameleon models explored in this work with the different concentration values (determined using the two different methods) and found them to be very similar. All the key figures given in the upcoming sections were based on the concentration values deduced from the maximum circular and virial velocities by the AHF algorithm. For comparison purposes we also present the results based on the concentration values deduced by directly fitting the AHF radial density profiles. Specifically, we fitted the density profiles of each cluster by allowing the $r_{200}$ to vary by $20 \%$ with respect to the value found by the halo finder software. This was found to improve the fit. The direct fit results along with the obtained distribution of the concentration parameter are described in detail in appendix \ref{direct_fit_results}.

Another important point to emphasize is that the used dataset was produced using standard cosmology simulations. One might argue that it might be incorrect to study the modified gravity effects using density distributions based on simulation data produced with an underlying assumption of gravity being described by GR. In other words, the existence of a fifth force would presumably affect structure formation and the resultant properties of cluster halos. Thus, in an ideal case, we would like to use halos from modified gravity simulations, which correctly account for the effects of the fifth force on structure formation and halo properties. However, we expect the fifth force effects to be very small and hence, they should not significantly alter cluster halos. Thus we do not expect the outlined bias coming from using GR simulations to significantly affect the results outlined in this work.

\begin{figure}
  \centering
    \includegraphics[width=0.6\textwidth]{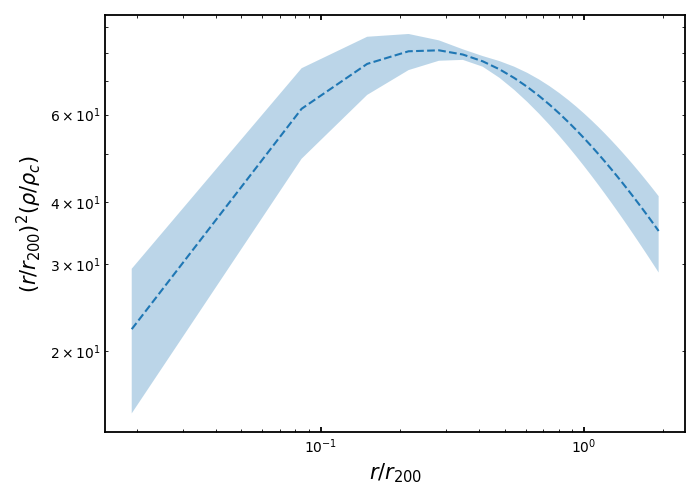}
    \caption{The density profiles calculated using the NFW parameters from the T{\scriptsize HE} T{\scriptsize HREE} H{\scriptsize UNDRED} P{\scriptsize ROJECT} dataset: the dashed line is the mean, while the light blue contour corresponds to the standard deviation. The data corresponds to the 317 cluster density profiles (7 clusters with invalid concentration values were removed) at redshift $z = 0$.}
    \label{project300_profiles}
\end{figure}

\begin{figure}
\centering
  \begin{subfigure}[b]{0.49\textwidth}
  \centering
    \includegraphics[width=0.99\textwidth]{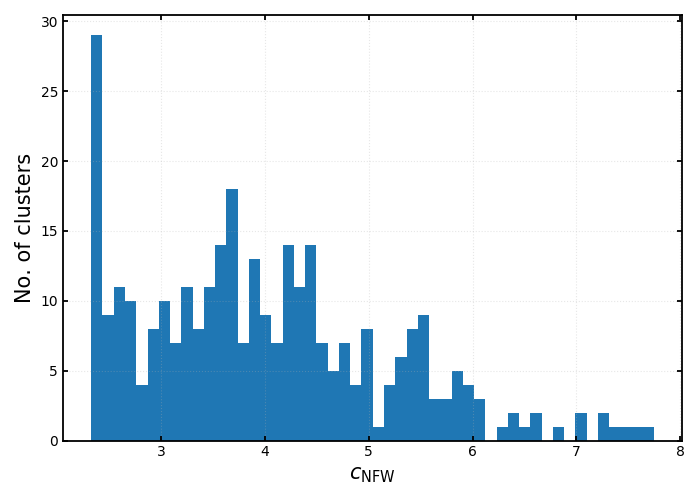}
    \label{project300:a1}
    
  \end{subfigure}
  \begin{subfigure}[b]{0.49\textwidth}
  \centering
    \includegraphics[width=0.99\textwidth]{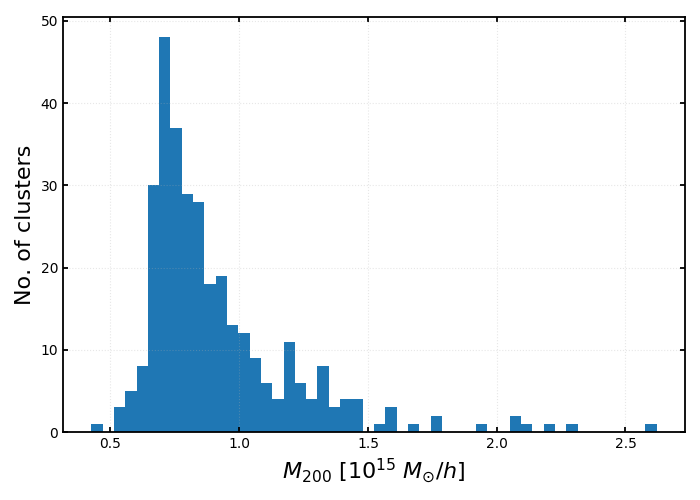}
    \label{project300:a2}
  \end{subfigure}
  \caption{The distribution of the concentration parameter, $c_{\rm NFW}$, and the corresponding $M_{200}$ values of the central clusters in the T{\scriptsize HE} T{\scriptsize HREE} H{\scriptsize UNDRED} P{\scriptsize ROJECT} dataset. }
  \label{project300_histograms}
\end{figure}

The dataset used in this work is summarized in figure \ref{project300_profiles}, which shows the NFW profiles produced by using the concentration and the $M_{200}$ parameters for each of the 324 central clusters at $z = 0$. Figure \ref{project300_histograms} illustrates the distribution of the concentration parameter and the corresponding virial mass. 

\section{The Finite Element Method Approach}
\label{FEM_introduction}

The FEM method refers to a widely used numerical technique for solving differential equations in one, two or three spatial dimensions. The technique has been traditionally used to solve differential equations in a wide array of contexts, namely problems in engineering, aerodynamics, fluid mechanics, meteorology and physics \cite{Zienkiewicz2013, Konrad1995, Ramirez1997}. The FEM method has also been recently applied for solving field equations of modified gravity models \cite{Burrage2018B, Braden2021, Burrage2021}.

FEM works by subdividing the problem domain into smaller subdomains known as finite elements. This is achieved by constructing a mesh with a finite number of points. The differential equation at hand is then rewritten in an integral form using Green's theorem, linearized if necessary and solved at each point of the mesh. More specifically, the domain $\Omega$ is discretized into a triangular mesh with boundaries of each element defined by their vertices $P_{i}$. The value of the variable of interest (e.g. the chameleon field $\phi$) can then be approximated as a piecewise polynomial function such that: 

\begin{equation}
\phi(\underline{x})=\sum_{i} U_{i} e_{i}(\underline{x}),
\end{equation}

\noindent with $U_{i} = U(P_{i})$ and $e_{i}$ as the basis functions defined such that $e_{i}\left(P_{j}\right)=\delta_{i j}$.

In our case we are interested in solving the chameleon field equation, eq.~(\ref{field_equation}), in the domain $\Omega$ with the boundary condition $\phi_{\rm BC} = \phi(\rho_{\rm BC})$ for some boundary density at the boundary of the domain $\partial \Omega$. The first step consists of rewriting the field equation using Green's theorem as follows:

\begin{equation}
\int_{\Omega}\left(\nabla^{2} \phi\right) v_{j} \mathrm{d} x+\int_{\Omega} \nabla \phi \cdot \nabla v_{j} \mathrm{d} x=\int_{\partial \Omega}\left(\partial_{n} \phi\right) v_{j} \mathrm{d} x,
\label{greens_theorem}
\end{equation}

\noindent where $v_{j}$ is known as the test function with the property that it vanishes on $\partial \Omega$ for all $j$ to satisfy the boundary conditions. Note that this property also means that the right hand side of eq.~(\ref{greens_theorem}) vanishes, and hence: 

\begin{equation}
\int_{\Omega} \nabla \phi \cdot \nabla v_{j} \mathrm{d} x=\int_{\Omega}\left(\nabla^{2} \phi\right) v_{j} \mathrm{d} x,
\label{greens_theorem2}
\end{equation}

\noindent where we can substitute eq.~(\ref{field_equation}) into the right hand side of eq.~(\ref{greens_theorem2}). 

Before proceeding it is useful to rescale the field equation in order to obtain a dimensionless form. In particular, we can rescale the field and the density by the corresponding background values: $\hat{\phi} = \phi/\phi_{\infty}$, $\hat{\rho} = \rho/\rho_{\infty}$. Similarly, the derivatives can be rescaled using the domain size $L$: $\hat{\nabla}^{2}=L^{2} \nabla^{2}$. This results in the following form for the rescaled field equation (note that this is the same rescaling that was applied in Ref.~\cite{Selkie}):

\begin{equation}
\alpha \hat{\nabla}^{2} \hat{\phi}=-\hat{\phi}^{-(n+1)}+\hat{\rho},
\label{rescaled_field_equation}
\end{equation}

\noindent with the $\alpha$ parameter defined as:

\begin{equation}
\alpha \equiv\left(\frac{M \Lambda}{L^{2} \rho_{\infty}}\right)\left(\frac{n M \Lambda^{3}}{ \rho_{\infty}}\right)^{\frac{1}{n+1}}.
\label{alpha}
\end{equation}

\noindent In order to linearize eq.~(\ref{rescaled_field_equation}) we can Taylor expand the non-linear term: 

\begin{equation}
\begin{aligned}
\hat{\phi}^{-(n+1)} & \approx \hat{\phi}_{k}^{-(n+1)}-(n+1) \hat{\phi}_{k}^{-(n+2)}\left(\hat{\phi}-\hat{\phi}_{k}\right)+\mathcal{O}\left(\hat{\phi}-\hat{\phi}_{k}\right)^{2} \\
\approx &(n+2) \hat{\phi}_{k}^{-(n+1)}-(n+1) \hat{\phi}_{k}^{-(n+2)} \hat{\phi}+\mathcal{O}\left(\hat{\phi}-\hat{\phi}_{k}\right)^{2} .
\end{aligned}
\label{taylor_expansion}
\end{equation}

\noindent Substituting eq.~(\ref{taylor_expansion}) into eq.~(\ref{rescaled_field_equation}) and then into eq.~(\ref{greens_theorem2}) we obtain the final form of the variational field equation (with higher order terms neglected): 

\begin{equation}
\alpha \int_{\Omega} \hat{\nabla} \hat{\phi} \cdot \hat{\nabla} v_{j} \mathrm{d} x+\int_{\Omega}(n+1) \hat{\phi}_{k}^{-(n+2)} \hat{\phi} v_{j} \mathrm{d} x=\int_{\Omega}(n+2) \hat{\phi}_{k}^{-(n+1)} v_{j} \mathrm{d} x-\int_{\Omega} \hat{\rho} v_{j} \mathrm{d} x.
\label{final_field_equation}
\end{equation}

\noindent Eq.~(\ref{final_field_equation}) can be solved iteratively by using the Picard iteration method, where we can use the k-th estimate of the field to deduce the k+1-th estimate: $\hat{\phi}_{k+1}=\omega \hat{\phi}+(1-\omega) \hat{\phi}_{k}$, with $\omega$ as the relaxation parameter \cite{Whiteman1994}.

In this work, in order to automate the solution of eq.~(\ref{final_field_equation}), we use SELCIE -- an open-source FEM solver based on the FEniCS Project software library \cite{Logg2010, Logg2012, Logg2012B, Selkie}. FEniCS refers to an open-source Python and C++ code optimized to solve linear and non-linear differential equations using a wide array of finite-element-based methods. FEniCS also contains the necessary tools to generate complex meshes in one, two and three dimensions along with a wide selection of linear and non-linear solvers as well as analysis tools for the obtained solutions.

SELCIE is optimized to solve the chameleon field equation on various scales. It automates the mesh creation process (by incorporating the \textit{Gmsh} package \cite{Geuzaine2009} or allowing the user to load their own mesh files), allows mesh refinement and produces solutions for 1D, 2D and 3D density distributions. SELCIE also allows us to model vacuum-chamber searches for the fifth force by solving the corresponding chameleon equation of motion. 

In order to solve eq.~(\ref{final_field_equation}) the following FEM approach was taken:

\begin{itemize}
    \item The problem domain was defined to be equal to $3 \times r_{200}$ for the spherical NFW halos (or correspondingly to $3 \times R_{200}$ in the triaxial case). This choice was motivated by trying to avoid the results being affected by the numerical artefacts that can manifest at the boundary of the domain and the regions with high density gradients ($\rho(r)$ goes to infinity at $r \rightarrow 0$).
    \item The mesh for the domain was generated using \textit{mshr} \cite{mshr}. Since the density distribution varies relatively slowly given the chosen domain size, no mesh refinement was used (note, however, that SELCIE allows mesh refinement by using \textit{Gmsh} if required). The FEniCS mesh precision setting was set to 200 (effectively, the number of mesh cells per unit length).      
    \item Since the NFW profile diverges at $r = 0$, to avoid unphysical densities, we introduced an inner cutoff at $r = r_{\rm cut}$ at which the density was set to a constant: $\rho_{\rm cut} = \rho_{}(r_{\rm cut})$. We explored different values of $r_{\rm cut}$, such that $r_{\rm cut} = \{0.03 \times r_{200}, 0.15 \times r_{200}, 0.45 \times r_{200} \}$. The middle value is a good estimate of a typical size of a cluster core, while the other two values were chosen to investigate the effects of a more extreme variation in the cutoff region sizes on the main results \cite{Sarazin1986,Ota2013}.
    \item An outer cutoff was introduced, such that if the density of a given cluster falls below the background density it is set to a constant: $\rho = \rho_{\infty}$. The background density was set to the critical density $\rho_{c} = 3H^{2}/8 \pi G$, with $H$ as the Hubble parameter. The background value of the density was then used to calculate the background field value $\phi_{\infty}$ using eq.~(\ref{phi_min}).
    \item The Picard iteration method code was run using SELCIE for a maximum of 300 iterations, using relaxation parameter; $\omega = 0.3$. The relaxation parameter was optimized by experimentation by taking into account that for high values of $\omega > 1$ the solver code might fail to converge, while for small values of $\omega < 0.01$ the convergence time becomes significantly longer. Finally, as a convergence criteria, we set $\delta \hat{\phi}= 10^{-14}$ as the solution tolerance limit, i.e. if the change in the field becomes smaller than the mentioned value, the code execution is stopped.
    \item For both spherical and triaxial halos we solve the chameleon equations in 2D. This is a natural choice for spherical halos, as there is no angular dependence in the chameleon equation of motion. In addition, producing 2D solutions requires less computational resources than their 3D equivalents. It should be noted that SELCIE allows solving problems in 1D, 2D and 3D \cite{Selkie}. For spherical density distributions a 3D problem can be reduced to a 2D problem. In the equation we are trying to solve, the volume integral can be replaced by the corresponding area element: $\mathrm{d}V \rightarrow \sigma \mathrm{d}A$, with $\sigma$ as the symmetry factor. The symmetry factors for an axis-symmetric system around the X-axis and Y-axis are given by $\sigma_{x} = |y|$ and $\sigma_{y} = |x|$. Throughout this work we use the vertical axis-symmetry ($\sigma = \sigma_{y}$). It should be noted that the results of this work are not affected by the choice of the symmetry factor.
    
    For the triaxial case, the studied halos are not spherically symmetric, however 2D solutions are sufficient to study the effects of interest (e.g. the angular dependence of the chameleon-to-NFW acceleration ratio as shown in figure \ref{triaxiality_effects}). Lastly, the lower-dimensional solutions are of special importance in the context of observational tests. More specifically, in real observational data we never observe the full 3D matter distribution, but rather work with a 2D projection on the sky. Hence, understanding the effects of triaxiality in 2D density profiles is of special interest in the context of the future observational searches for fifth forces.    
\end{itemize}

\begin{figure}
  \centering
    \includegraphics[width=0.90\textwidth]{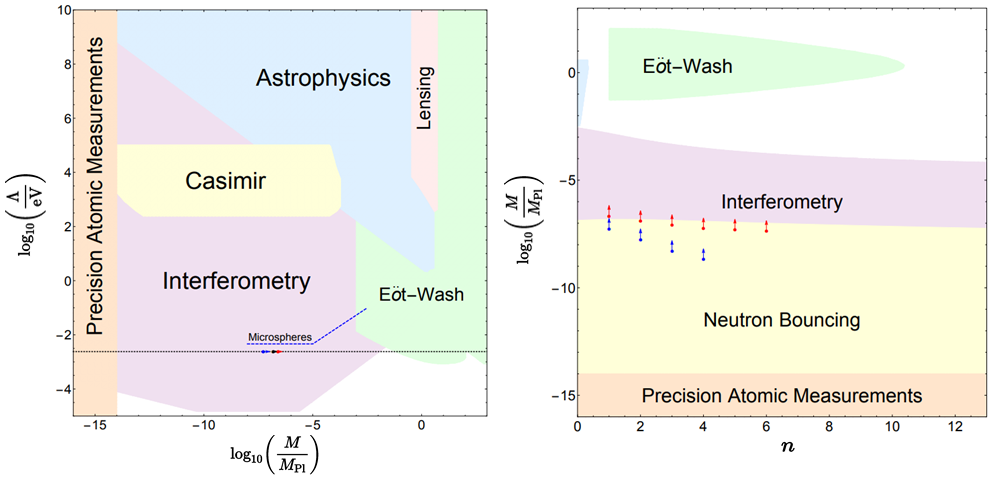}
    \caption{Observational and laboratory constraints of chameleon gravity. The regions excluded by each specific test are indicated in the figure. The figure on the left shows the constraints on $\Lambda$ and $M$, for $n = 1$. Figure on the right shows the corresponding constraints for $n$ and $M$ for $\Lambda = 2.4$ meV. The region labelled astrophysics contains the bounds from both Cepheid and rotation curve tests. The dashed line indicates the dark energy scale $\Lambda = 2.4$ meV. The black, red, and blue arrows in both figures show the lower bound on $M$ coming from neutron bouncing and interferometry experiments. In this work we focus on the white regions of the parameter space. Figure adapted from \cite{Burrage2018}. }
    \label{chameleon_constraints}
\end{figure}

The code was run using the outlined settings for each cluster in our dataset. Once the solution was obtained, the following quantities were calculated: the field profile and the gradient over the entire domain, the NFW mass distribution, the NFW potential as well as the relative chameleon and NFW acceleration profiles. In order to investigate the effects of triaxiality, the ratio of the chameleon and the NFW acceleration at $r_{200}$\footnote{Note that we cannot use the triaxial radius $R_{200}$ here, as it has an angular dependence such that it would be gauging the same exact density value along the X and the Y axes. Otherwise, any distance scale could be used, so we chose $r_{200}$ as it simplifies the calculations.} was calculated along the X and the Y axes. Finally, the results were compared against the analytical approximation of the solution. The further settings of the code related to the studied chameleon models, the analytical tests of the solutions and the feature importance study are listed below:

\begin{itemize}
    \item The chameleon model parameters, i.e. $\Lambda$, $M$ and $n$, were chosen from the regions that are still currently allowed by the available observational constraints (see figure \ref{chameleon_constraints} and Ref.~\cite{Burrage2018} for a detailed analysis on the observational constraints). More specifically, we explore $\Lambda$ values that are close to the dark energy scale of $2.4$ meV. Current constraints have almost entirely ruled out the chameleon models with $\Lambda = 2.4$ meV, hence we explore lower values in the region of $\Lambda \in [10^{-3}, 10^{-5}]$ eV along with the corresponding coupling parameters, such that $M \in [10^{-5},10^{3}] \times M_{\rm Pl}$ and $M \in [10^{-14}, 10^{-10}] \times M_{\rm Pl}$. The mentioned parameter values correspond to the $n = 1$ model, however, we can also allow $n$ to vary. To investigate what effects $n$ has on the results, we run the code for a number of selected values of $n$, such that $n \in [1,12]$. These runs are produced with $M = 10^{-3}- 10^{-4}\times M_{\rm Pl}$ as we require a small value of $M$ for the chameleon effects to be significant and so we choose the smallest value currently allowed by the observational constraints (for $n \neq 1$). Since we found $M$ to be the key parameter in obtaining the largest values of the chameleon-to-NFW acceleration ratios allowed by the observational constraints, in our analysis we primarily focused on the parameter space region with $M \in [10^{-14},10^{-11}]\times M_{\rm Pl}$ and $\Lambda \in [10^{-4}, 10^{-5}]$ eV.

    \item As a first test the results of the FEM numerical solver were compared against the existing analytic solutions. For instance, for a constant density spherical source, there exists a known approximate solution for a field profile given by the following \cite{Khoury2004}\footnote{Note that the equation given here is the rescaled version of the analytic solution presented in Ref.~\cite{Khoury2004}.}:

    \begin{equation}
        \hat{\phi}_{\rm sph} (\hat{r}) = 1 - \frac{\hat{R}_{\rm src}}{\hat{r}} \exp\bigg[-\bigg(\hat{r} - \hat{R}_{\rm src} \bigg) \sqrt{\frac{n + 1}{\alpha}}   \bigg],
        \label{sph_solution}
    \end{equation}
    
    \noindent with $\hat{R}_{\rm src}$ as the spherical source radius, $\hat{r}$ as the radial coordinate (both quantities rescaled by the domain size $L$) and $\alpha$ given by eq.~(\ref{alpha}). Similarly, given our model parameter choice and the size of a typical galaxy cluster, the obtained $\alpha$ values are very small. This means that we expect the obtained numerical solution to be well approximated by eq.~(\ref{phi_min}), or, in rescaled units, by:
    
    \begin{equation}
    \hat{\phi}_{\min } =\hat{\rho}^{-\frac{1}{n+1}}.
    \label{phi_min_dimensionless}
    \end{equation}

    \noindent As an extra consistency check the numerical results were compared against the approximation given by the equation above. 
    
    \item The triaxial results were further studied in order to determine the feature importance. More specifically, in order to understand whether the virial mass $M_{200}$, the concentration $c_{\rm NFW}$ or the shape parameters $q_{a}$ and $q_{c}$ have a higher influence on the chameleon acceleration at the triaxial virial radius, we studied the importance of each of the aforementioned variables. This was done by using three different algorithms: linear regression, decision trees and gradient boosted decision trees (i.e. the XGBoost algorithm \cite{Chen2016}). These three machine learning algorithms offer a natural way of inferring which of the input features are the most important when determining the output variable (chameleon acceleration or the chameleon-to-NFW acceleration ratio in our case). More specifically, we used the acceleration ratio values for each cluster in our dataset as the target variable, which we aimed to predict using the features. In the case of linear regression, the regression coefficients (assuming the data is correctly normalized) are directly related to feature importance. In the case of decision trees (including the gradient boosted trees), the feature importance is generally calculated by evaluating the Gini importance (here given for a simple case of only two child nodes) \cite{Menze2009}: 

\begin{equation}
    G_{j} = w_{j}C_{j} - w^{L}_{j}C^{L}_{j} - w^{R}_{j}C^{R}_{j},
\end{equation}
    
    \noindent with $G_{j}$ as the importance of node $j$ and $w_{j}$ as the weighted number of samples reaching node $j$. The $w^{L}_{j}$, $w^{R}_{j}$, $C^{L}_{j}$ and $C^{R}_{j}$ terms correspond to the number of samples in the child node from the left/right split on node $j$ and the corresponding impurity value of the child node. The importance of feature $i$ is then calculated by:
    
    \begin{equation}
        I_{i}  = \frac{\sum_{j}^{s} G_{j}}{\sum_{k}^{N} G_{k}},
    \end{equation}
     
     \noindent with $s$ as the number of node $j$ splits for feature $i$, and $N$ as the total number of nodes in the decision tree.

\end{itemize}

\section{Spherical NFW Halo Results}
\label{spherical_results}

\subsection{Testing the Solver}

\begin{figure}
\centering
  \begin{subfigure}[b]{0.49\textwidth}
  \centering
    \includegraphics[width=0.99\textwidth]{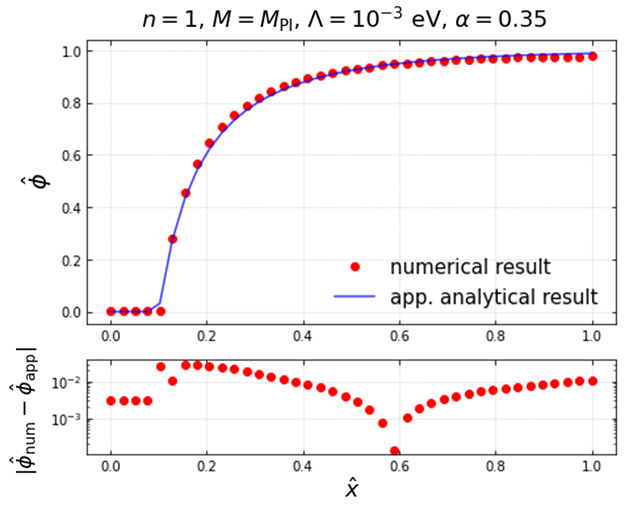}
    \label{fig4:a1}
    
  \end{subfigure}
  \begin{subfigure}[b]{0.48\textwidth}
  \centering
    \includegraphics[width=0.99\textwidth]{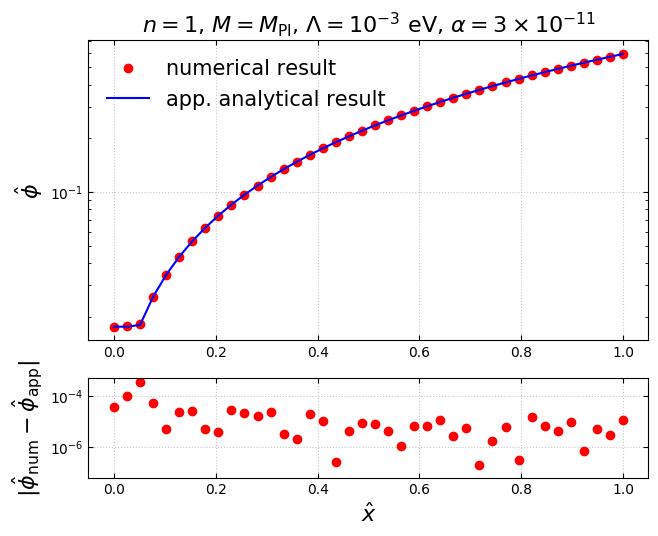}
    \label{fig4:a2}
  \end{subfigure}
  \caption{A comparison between the numerical and approximate analytical solutions for small and larger values of $\alpha$. \textit{Left}: comparing the numerical solution against the analytic solution for a spherical source of constant density for $\alpha = 0.35$. \textit{Right}: a comparison between the numerical results and the analytic approximation for $\alpha = 1 \times 10^{-11}$ for a typical cluster in our dataset.   }
  \label{numeric_vs_analytic}
\end{figure}

To check the validity of the solutions, the FEM solver results were compared against the known analytical solution, eq.~(\ref{sph_solution}), for relatively larger and smaller values of the $\alpha$ parameter. Figure \ref{numeric_vs_analytic} shows the comparison between the numerical results and the approximate analytical solution for a spherical source (with constant density) with $\hat{R}_{\rm src} = 0.1$, $\hat{\rho} = 10^{5}$, $\alpha = 0.35$ and $n = 1$. The figure shows good agreement between the numerical results and the analytic approximation, with sub-percent level residuals throughout most of the domain with exception for the region where the field profile varies rapidly and the residuals can reach 5-10$\%$ level. Note, however, that the analytic approximation is not the exact solution to the field equation and hence we naturally expect to see some difference between the numeric and the analytic results in the region where the field changes rapidly (e.g. see the recent discussion on chameleon field properties for point particles in Ref.~\cite{Burrage2021}). 

Figure \ref{numeric_vs_analytic} also shows the comparison between the numerical results and the analytic approximation for a typical galaxy cluster in our dataset for $\alpha \ll 1$. In this regime, the field profile can be approximated well by eq.~(\ref{phi_min_dimensionless}). We found that the agreement with the approximate solution along with the residuals were nearly identical for all of the clusters in our dataset.

\begin{figure}
\centering
  \begin{subfigure}[b]{0.49\textwidth}
  \centering
    \includegraphics[width=0.9\textwidth]{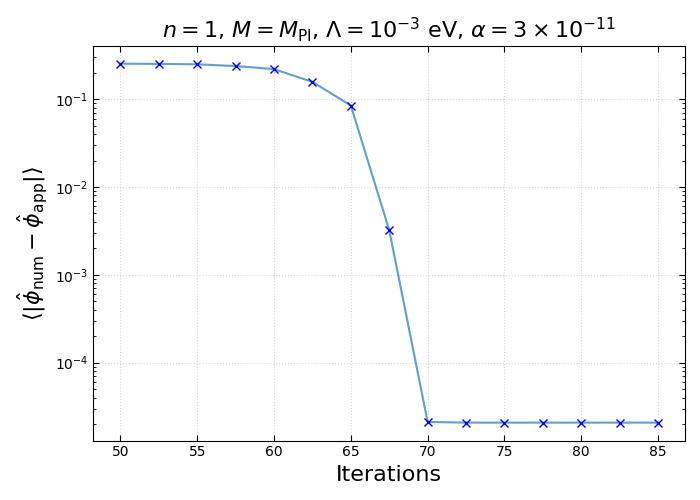}
    \label{fig5:a1}
    
  \end{subfigure}
  \begin{subfigure}[b]{0.49\textwidth}
  \centering
    \includegraphics[width=0.90\textwidth]{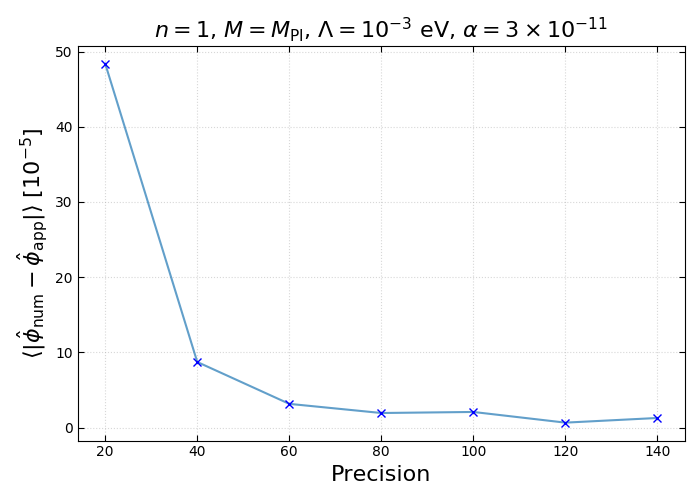}
    \label{fig5:a2}
  \end{subfigure}
  \caption{The value of the mean absolute residuals for a different number of iterations of the code and the different values of the mesh precision for a typical cluster in our dataset. }
  \label{residual_analysis}
\end{figure}

The residuals were further studied by varying the different parameters in our model. Generally we found that there is no significant relation between the size of the residuals and $\Lambda$ or $M$. There is some dependence on $n$, however, this is expected, as different values of $n$ correspond to different chameleon models, with generally different values of the field. There is, however, a significant dependence on the value of the mesh precision (effectively, the mesh resolution). Figure \ref{residual_analysis} illustrates the mean residual dependence on the mesh precision. Figure \ref{residual_analysis} also illustrates that it takes the code around 70 iterations to converge (with $\delta \hat{\phi} = 10^{-14}$ as the convergence limit), after which the residuals remain effectively constant. 

\subsection{Spherical NFW Halo Results for Varying $M$}

For each NFW halo we calculated the chameleon-to-NFW acceleration ratio and its radial dependence. These are key quantities as they summarize how significant the chameleon effects are compared to the standard gravitational effects in the different parts of the cluster. Also, it is useful to define an effective chameleon mass $M_{\phi}$\footnote{Note that, more specifically, this refers to the mass sourcing the fifth force and is also known as the \textit{thin-shell} mass.}, by noting that $a_{\phi} = - GM_{\phi}/r^{2}$. The $a_{\phi}/a_{\rm NFW}$ ratio is then exactly equal to the ratio of the chameleon-to-NFW inertial masses, $M_{\phi}/M_{\rm NFW}$. This ratio can be compared against the currently available mass measurements on galaxy cluster scales to estimate whether the deduced chameleon effects would be measurable (see section \ref{section:conclusions} for a further discussion).  

\begin{figure}
\centering
  \begin{subfigure}[b]{0.49\textwidth}
  \centering
    \includegraphics[width=0.95\textwidth]{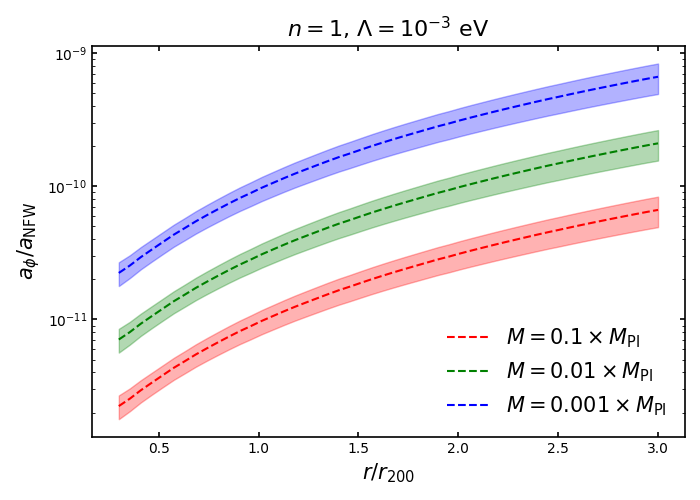}
    \label{fig6:a1}
    
  \end{subfigure}
  \begin{subfigure}[b]{0.49\textwidth}
  \centering
    \includegraphics[width=0.95\textwidth]{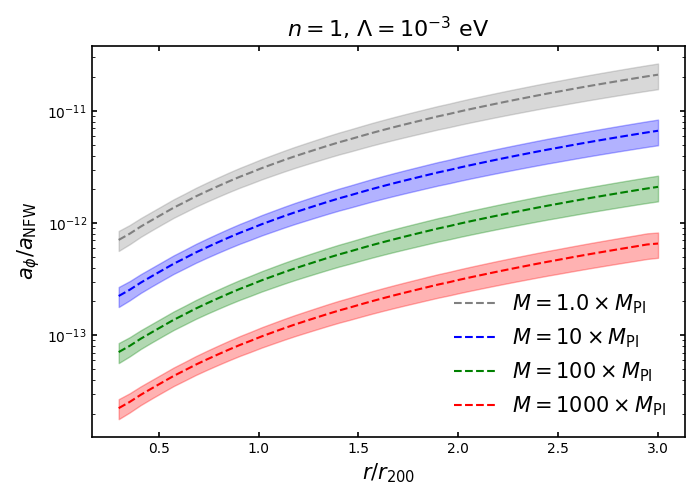}
    \label{fig6:a2}
  \end{subfigure}
  \caption{The results for the chameleon-to-NFW acceleration ratio for the full cluster dataset for different $M$ (coupling) values. The dashed lines correspond to the mean profile, while the coloured contours correspond to the standard deviation.}
  \label{main_results_vs_M}
\end{figure}

Figure \ref{main_results_vs_M} illustrates the variation of the $a_{\phi}/a_{\rm NFW}$ profile for different $M$ parameter values. The results show that for $M$ between $10^{-3}-10^{3} \times M_{\rm Pl}$, the ratio $a_{\phi}/a_{\rm NFW}$ is in the range of around $10^{-13}-10^{-10}$ when measured at $r_{200}$. An important conclusion is that to get the highest possible value of $a_{\phi}/a_{\rm NFW}$, the smallest possible value of $M$ allowed by the observational constraints is required. However, even for the smallest allowed value of $M$, the acceleration ratio is likely too small to be cosmologically significant.

\begin{figure}
  \centering
    \includegraphics[width=0.75\textwidth]{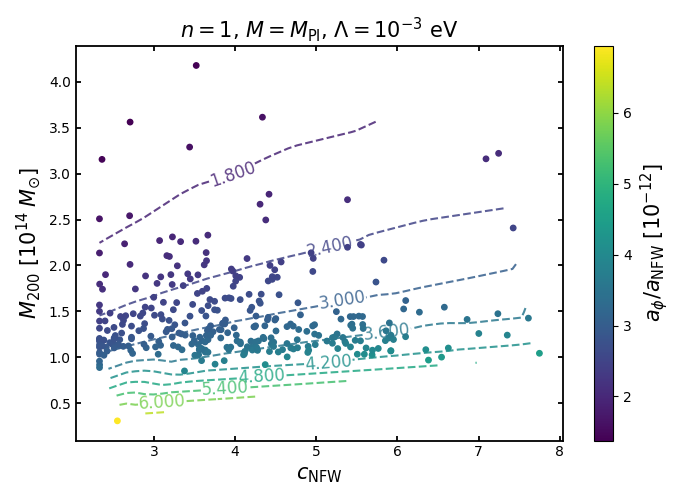}
    \caption{The chameleon-to-NFW acceleration ratios measured at the virial radius for the full cluster dataset. Each point represents an individual cluster, with the colour scale based on the $a_{\phi}/a_{\rm NFW}$ value. The dashed contours represent lines of equal $a_{\phi}/a_{\rm NFW}$.}
    \label{contour_plot_M1E0_cutoff_0p05}
\end{figure}

Figure \ref{contour_plot_M1E0_cutoff_0p05} illustrates the $a_{\phi}/a_{\rm NFW}$ results, as measured at $r_{200}$, for each individual cluster in the dataset. The results illustrate that there is a non-trivial relation between the NFW parameters and the acceleration ratio. More specifically, for constant $M_{200}$, the acceleration ratio is larger for higher concentrations, $c_{\rm NFW}$. Also, as illustrated by the outlier point (with $c_{\rm NFW} \approx 2.5$ and $M_{200} \approx 2.5 \times 10^{13} \; M_{\odot}$), for very low masses and low concentrations, the acceleration ratio is significantly higher (the chameleon acceleration is still extremely small compared to the corresponding NFW acceleration). The dashed lines show the contours of equal acceleration ratio values, which have been extrapolated to areas with a lower number of points. It should be noted that in regions with very few points, the contour values might not be accurate.    

\begin{figure}
\centering
  \begin{subfigure}[b]{0.49\textwidth}
  \centering
    \includegraphics[width=1.10\textwidth]{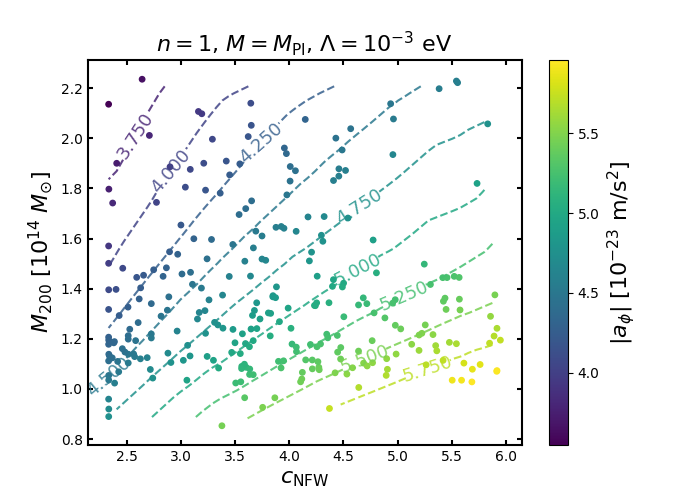}
    \label{main_results:a1}
    
  \end{subfigure}
  \begin{subfigure}[b]{0.49\textwidth}
  \centering
    \includegraphics[width=1.10\textwidth]{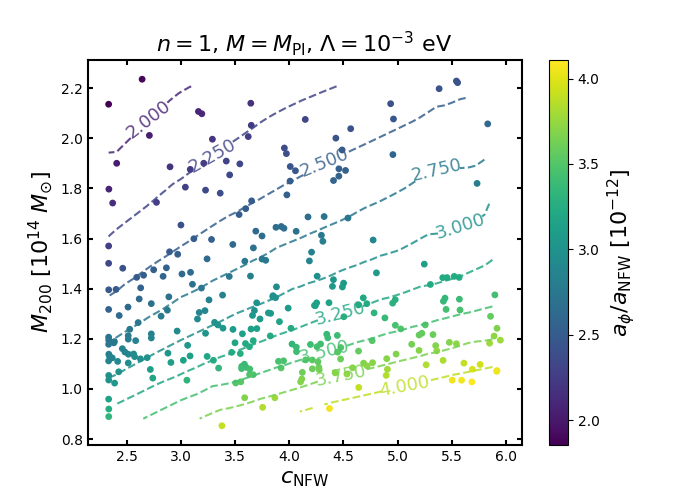}
    \label{main_results:a2}
  \end{subfigure}
  \caption{\textit{Left:} the chameleon acceleration results with outlier points removed to allow more accurate calculations of the contour lines. \textit{Right:} the acceleration ratio results with the outlier points removed.}
  \label{main_results_no_outliers}
\end{figure}

Figure \ref{main_results_no_outliers} shows the results for the chameleon-to-NFW acceleration ratio, $a_{\phi}/a_{\rm NFW}$, zoomed in on the most densely populated region of the parameter space. Specifically, we show the results for $0.8 \times 10^{14} \; M_{\odot} < M_{200} < 2.2 \times 10^{14} \; M_{\odot}$ and $c_{\rm NFW} < 6.0$. This allows us to calculate the equal-value contour lines more accurately. The results show a complicated relationship between the NFW parameters and the chameleon acceleration. Generally, given our dataset, smaller values of the virial mass along with high concentration lead to higher chameleon acceleration when measured at the virial radius. The same conclusion can be drawn for the chameleon-to-NFW acceleration ratio. The relationship between the concentration and the size of the chameleon acceleration shown in figure \ref{main_results_no_outliers} might seem counter-intuitive as one might expect the chameleon effects being more screened for higher concentrations at constant $M_{200}$. However, as we discuss in appendix \ref{appendix_analytical_approx}, the relationship between $c_{\rm NFW}$ and $a_{\phi}/a_{\rm NFW}$ is complicated and depends on the radial distance from the cluster center at which the acceleration measurements are made.

\begin{figure}
\centering
\makebox[\linewidth][c]{%
  \begin{subfigure}[b]{0.37\textwidth}
  \centering
    \includegraphics[width=1.00\textwidth]{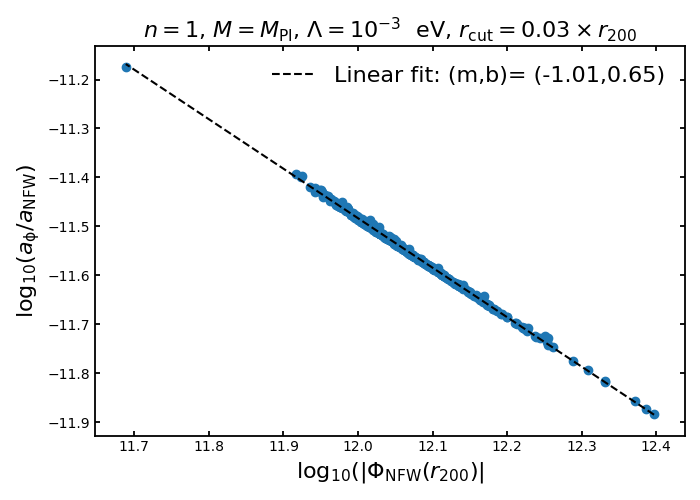}
    \label{NFW_potential_vs_acceleration_1}
    
  \end{subfigure}
  \begin{subfigure}[b]{0.37\textwidth}
  \centering
    \includegraphics[width=1.00\textwidth]{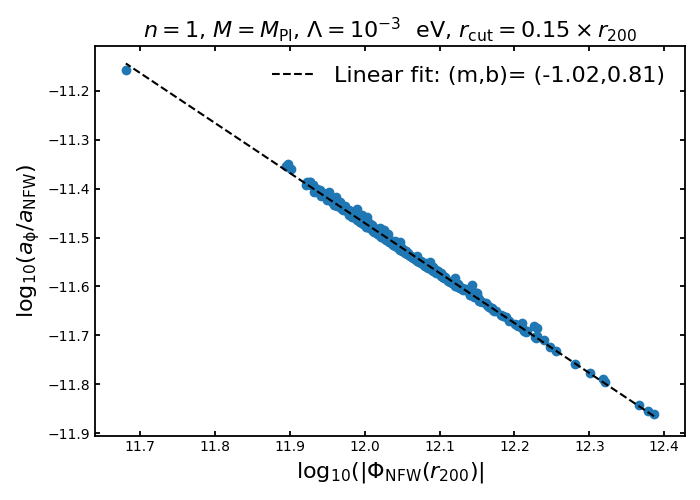}
    \label{NFW_potential_vs_acceleration_2}
  \end{subfigure}
  \begin{subfigure}[b]{0.37\textwidth}
  \centering
      \includegraphics[width=1.00\textwidth]{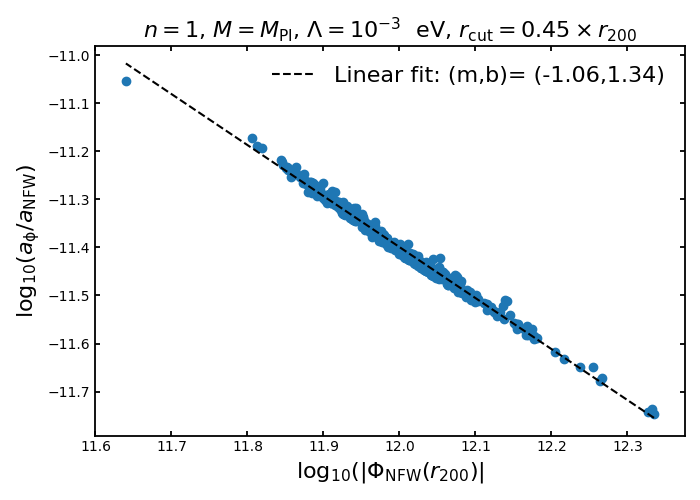}
      \label{NFW_potential_vs_acceleration_3}
  \end{subfigure}
  }
  \caption{The relationship between the NFW potential and the chameleon-to-NFW acceleration ratio at the virial radius for different sizes of the inner cutoff radius (size of the core) for the NFW halo. The dashed lines show the results of the linear fit, with $m$ as the slope and $b$ as the intercept parameter.}
  \label{NFW_potential_vs_acceleration}
\end{figure}

Figure \ref{NFW_potential_vs_acceleration} illustrates that even though the relationship between the acceleration ratio and individual NFW parameters is complicated, there exists a straightforward relationship between the NFW potential and the chameleon acceleration. Namely, for smaller values of the potential, there is less screening, which leads to a higher acceleration ratio. This agrees well with the intuition established in the literature (e.g. see section 3 in \cite{Burrage2016}). The results also agree well with the relationship between the Newtonian potential and the screening factor established in the literature \cite{Burrage2018}. Figure \ref{NFW_potential_vs_acceleration} also illustrates that the scatter of the points depends primarily on the $r_{\rm cut}$ value. In other words, there is a subtle relation between the size of the cluster core and the magnitude of chameleon screening. The size of the inner cutoff region changes both the scatter and the slope of the fit.

\subsection{Spherical NFW Halo Results for Varying $n$}

\begin{figure}
\centering
\makebox[\linewidth][c]{%
  \begin{subfigure}[b]{0.49\textwidth}
  \centering
    \includegraphics[width=1.00\textwidth]{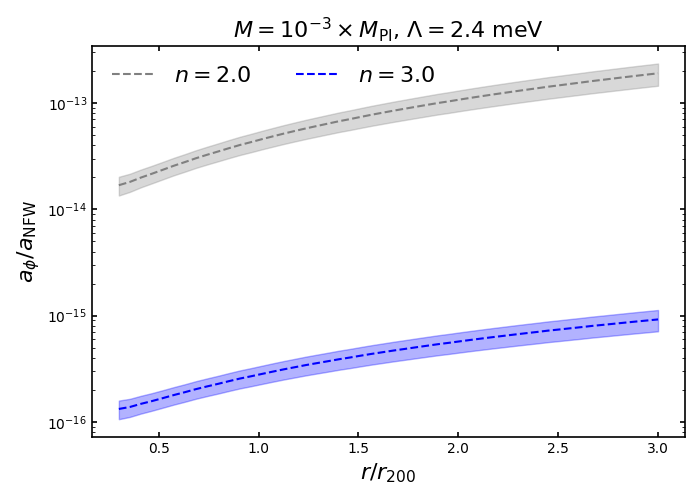}
    \label{a_ratio_vs_varying_n:a1}
    
  \end{subfigure}
  \begin{subfigure}[b]{0.49\textwidth}
  \centering
    \includegraphics[width=1.00\textwidth]{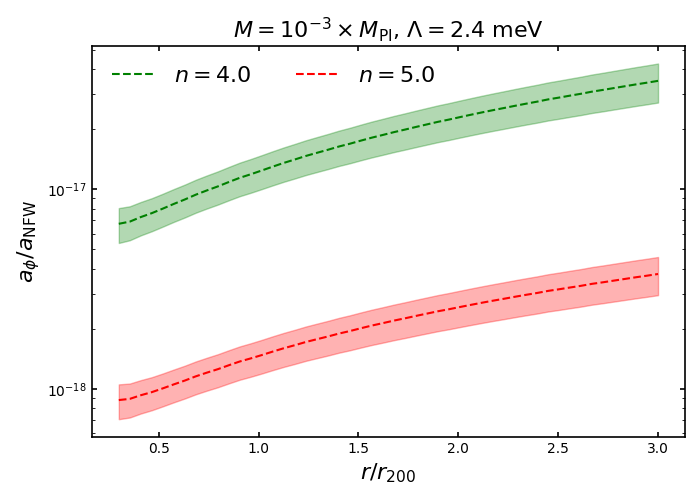}
    \label{a_ratio_vs_varying_n:a2}
  \end{subfigure}
  }
  \caption{The chameleon-to-NFW acceleration ratio for the different $n$ values. The dashed lines correspond to the mean profile, while the coloured contours correspond to the standard deviation.}
  \label{a_ratio_vs_varying_n}
\end{figure}

In addition to varying the coupling parameter $M$, we also investigated the effects of the different chameleon models corresponding to the different values of the $n$ parameter. Specifically, we focused on the models with positive $n$ values, which are less constrained than the corresponding negative $n$ models and possess interesting phenomenology in the context of quintessence \cite{Copeland2006}. It was found that increasing the value of $n$ results in a significant reduction of the $a_{\phi}/a_{\rm NFW}$ ratio (when compared to the $n = 1$ case). The results for the $a_{\phi}/a_{\rm NFW}$ are summarized in figure \ref{a_ratio_vs_varying_n}.

\begin{figure}
\centering
  \begin{subfigure}[b]{0.49\textwidth}
  \centering
    \includegraphics[width=1.00\textwidth]{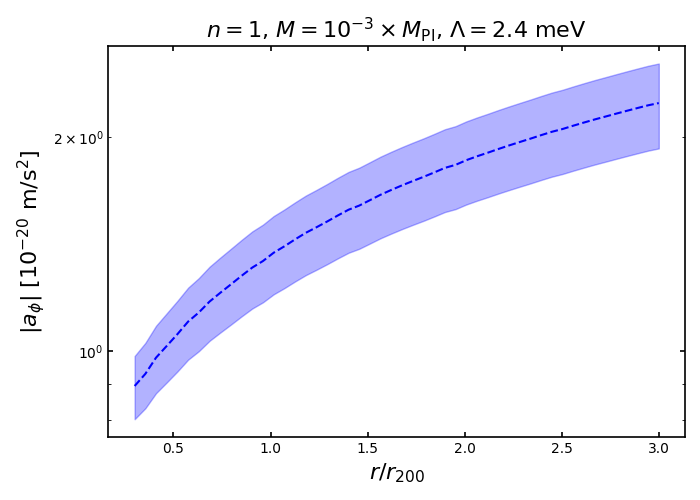}
    \label{a_phi_n_1_n_3:a1}
    
  \end{subfigure}
  \begin{subfigure}[b]{0.49\textwidth}
  \centering
    \includegraphics[width=1.00\textwidth]{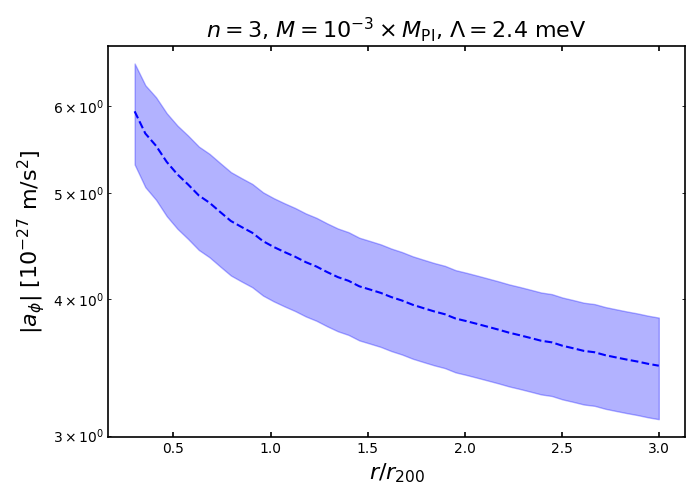}
    \label{a_phi_var_n_1_n_3:a2}
  \end{subfigure}
  \caption{The chameleon acceleration results for $n = 1$ and $n = 3$.}
  \label{chameleon_acceleration_n}
\end{figure}

Another key difference between the chameleon models (i.e. models with $n = 1$ and $n > 1$) is that the radial profile of $a_{\phi}(r)$ is changed significantly. For $n = 1$, $a_{\phi}(r)$ increases radially, while in the $n > 1$ case $a_{\phi}(r)$ grows towards the centre. However, it should be noted that $a_{\phi}/a_{\rm NFW}$ still decreases radially, as $a_{\rm NFW}$ becomes smaller at a relatively quicker rate (i.e. the gradient of the $a_{\rm NFW}(r)$ curve is steeper). Hence, the highest deviation from the Newtonian acceleration can be obtained in the outer regions of the cluster rather than the cluster core region. These results are summarized in figure \ref{chameleon_acceleration_n}, which shows the key differences in the chameleon acceleration profile for two different values of $n$.

\subsection{Spherical NFW Halo Results for Varying $\Lambda$}

\begin{figure}
\centering
  \begin{subfigure}[b]{0.49\textwidth}
  \centering
    \includegraphics[width=1.00\textwidth]{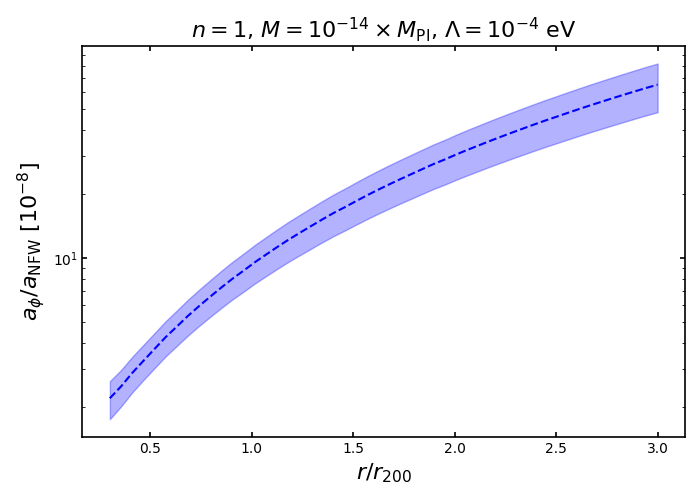}
    \label{a_ratio_small_lambda:a1}
    
  \end{subfigure}
  \begin{subfigure}[b]{0.49\textwidth}
  \centering
    \includegraphics[width=1.00\textwidth]{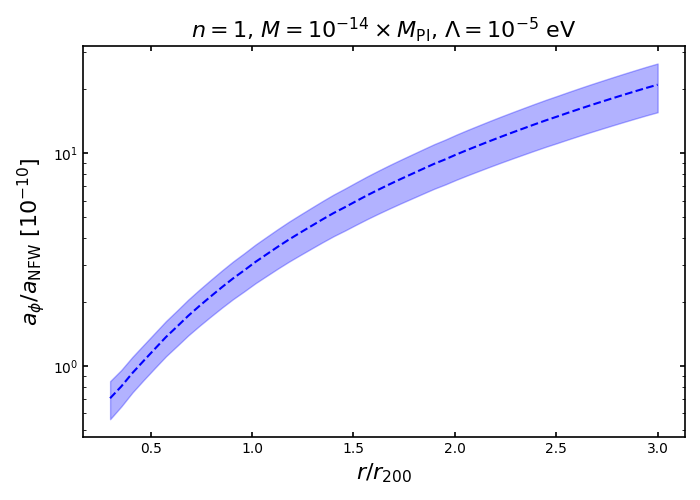}
    \label{a_ratio_small_lambda:a2}
  \end{subfigure}
  \caption{The chameleon-to-NFW acceleration ratio profiles for the varying values of the energy scale $\Lambda$.}
  \label{acceleration_ratios_small_Lambda}
\end{figure}

\begin{figure}
\centering
  \begin{subfigure}[b]{0.49\textwidth}
  \centering
    \includegraphics[width=1.10\textwidth]{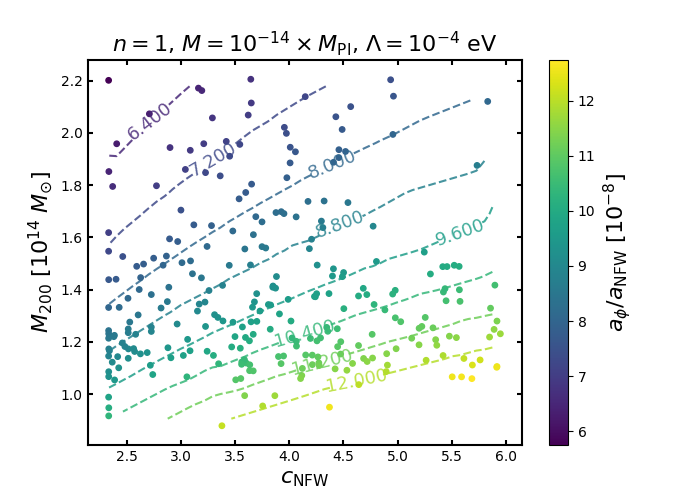}
    \label{main_results_lambda:a1}
    
  \end{subfigure}
  \begin{subfigure}[b]{0.49\textwidth}
  \centering
    \includegraphics[width=1.10\textwidth]{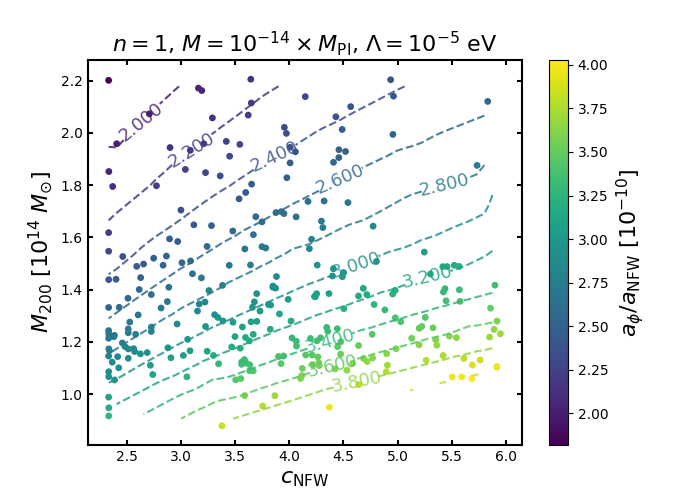}
    \label{main_results_lambda:a2}
  \end{subfigure}
  \caption{The chameleon-to-NFW acceleration ratio results vs. the NFW parameters for different values of $\Lambda$ (with outlier values removed).}
  \label{main_results_no_outliers_lambda}
\end{figure}

\begin{figure}
\centering
  \begin{subfigure}[b]{0.49\textwidth}
  \centering
    \includegraphics[width=1.00\textwidth]{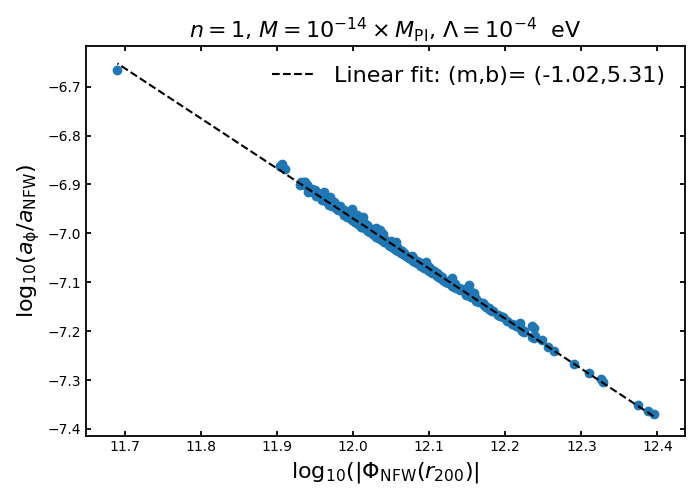}
    \label{NFW_pot_lambda:a1}
    
  \end{subfigure}
  \begin{subfigure}[b]{0.49\textwidth}
  \centering
    \includegraphics[width=1.00\textwidth]{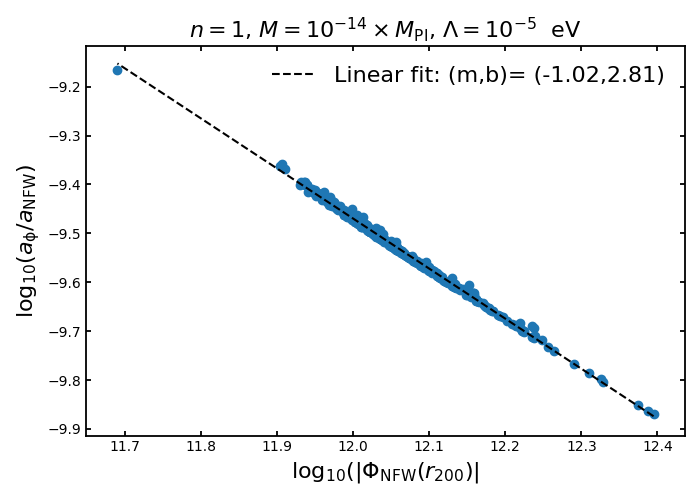}
    \label{NFW_pot_lambda:a2}
  \end{subfigure}
  \caption{The relationship between the NFW potential and the chameleon-to-NFW acceleration ratio for different values of $\Lambda$. }
  \label{acc_ratio_vs_NFW_potential_Lambdas}
\end{figure}

In addition to varying the values of the $M$ and $n$ parameters, we also investigated the effects of varying the energy scale $\Lambda$. As previously discussed, the case of $\Lambda$ being equal to the dark energy scale of 2.4 meV is highly constrained by current astrophysical and laboratory measurements. However, moving away from the dark energy scale, there exist models with interesting phenomenology that are not yet ruled out by the observational data. In this context, we were particularly interested in parameter values of $\Lambda \leq 10^{-4}$ eV and $M \approx 10^{-14} \times M_{\rm Pl}$ (as these parameters lead to the highest acceleration ratio values). The chameleon-to-NFW acceleration profiles for these choices of parameters are summarized in figure \ref{acceleration_ratios_small_Lambda}. The individual cluster results are shown in figure \ref{main_results_no_outliers_lambda}. Finally, the relation between the NFW potential and the acceleration ratio is given in figure \ref{acc_ratio_vs_NFW_potential_Lambdas}.

We found that the relationship between the NFW parameters and the chameleon-to-NFW acceleration remains nearly identical when compared to the other models mentioned in this work (i.e lower values of $M_{200}$ and higher concentrations $c_{\rm NFW}$ lead to higher $a_{\rm \phi}/a_{\rm NFW}$). Similarly, the radial $a_{\phi}/a_{\rm NFW}$ profile generally retains a similar shape when compared to the results in the previous sections. Lastly, the tight relation between the NFW potential and the chameleon-to-NFW ratio observed in the previous sections is also observed here. 

\section{The Effects of Triaxiality}
\label{triaxiality_results} 

\subsection{Full Triaxiality Results}
\label{full_triaxiality_results}

The results in this section illustrate the effects of the halo shapes on the acceleration ratios. More specifically, here we use the same dataset from the T{\scriptsize HE} T{\scriptsize HREE} H{\scriptsize UNDRED} P{\scriptsize ROJECT}, but now take into account the shapes as described by the AHF triaxiality parameters. It is important to note that the NFW concentration and virial mass that we use here were derived with an underlying assumption of spherical symmetry (i.e. they were derived from spherical density shells). In an ideal case, one would instead use a triaxial density profile to derive the mentioned parameters; however, a full dataset based on the triaxial NFW profiles was not available. Nonetheless, we do not expect the results to be affected significantly by this underlying assumption of spherical symmetry. For instance, as shown in Ref.~\cite{Knebe2006}, assuming spherical symmetry for a triaxial halo leads to errors of around 10-20\% when determining the density distribution for average ellipticity halos. We do not expect that a bias of such magnitude will significantly affect the conclusions drawn in this work.    

\begin{figure}
\centering
  \begin{subfigure}[b]{0.49\textwidth}
  \centering
    \includegraphics[width=1.00\textwidth]{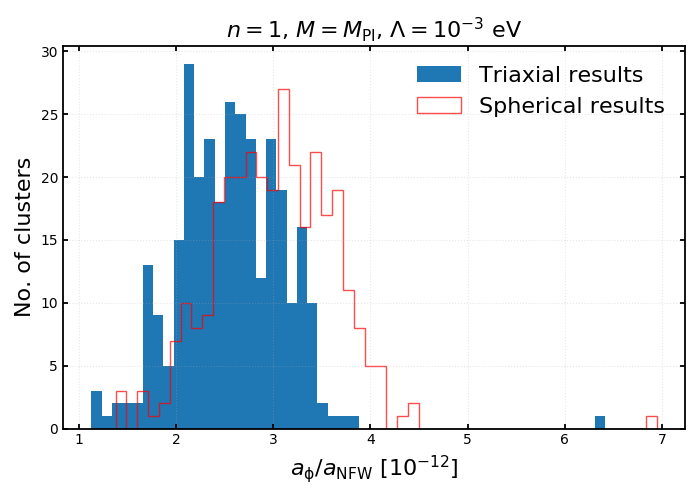}
    \label{main_results_triaxiality:a1}
    
  \end{subfigure}
  \begin{subfigure}[b]{0.49\textwidth}
  \centering
    \includegraphics[width=1.00\textwidth]{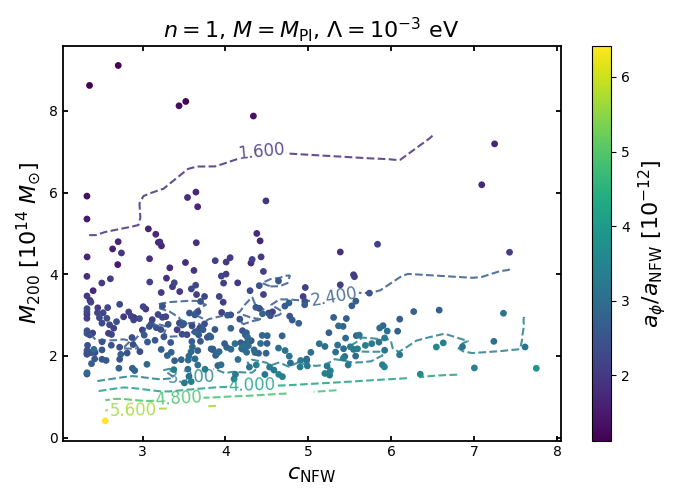}
    \label{main_results_triaxiality:a2}
  \end{subfigure}
  \caption{The chameleon-to-NFW acceleration ratios for triaxial halos. \textit{Left:} the comparison of the triaxial results (measured at $R_{200} = (q_{a}q_{b})^{-1/3} r_{200}$) against the spherical results (measured at $r_{200}$). \textit{Right:} the chameleon-to-NFW acceleration ratio at the triaxial virial radius vs. the NFW parameters. The contour lines show a degeneracy between the NFW parameters and the axis ratios.}
  \label{triaxial_vs_spherical_results}
\end{figure}

As described in section \ref{nfw_halos}, we also expect the effects of triaxiality on the chameleon-to-NFW acceleration ratio to be subtle. Namely, since the key calculations are now done at the triaxial virial radius (and generally we expect the axis ratios to be relatively close to unity), we expect the main results to be similar to the spherical case, but rescaled by some factor that depends on the X and Y axis ratios. In addition, we expect further subtle effects due to the dependence of screening effects on the source shape (as described in Ref.~\cite{Burrage2018B}). Figure \ref{triaxial_vs_spherical_results} illustrates the magnitude of the difference between the spherical and triaxial results. It clearly shows that the triaxial acceleration ratio distribution is nearly identical to the spherical one, albeit shifted towards slightly smaller values.

The relationship between the NFW parameters and the acceleration ratio, however, is now much more complicated as shown in figure \ref{triaxial_vs_spherical_results}. The general distribution of the points looks similar to the one observed in figure \ref{contour_plot_M1E0_cutoff_0p05}, however, the $a_{\phi}/a_{\rm NFW}$ values appear to be relatively more scattered (as illustrated by the contour lines). These results illustrate the non-trivial relationship between the NFW halo properties and their shape. Specifically, one can imagine two halos with the exact same NFW parameters, but radically different shapes. These two halos would result in different $a_{\phi}/a_{\rm NFW}$ values at the triaxial virial radius (and hence two differently coloured points at the same position in figure \ref{triaxial_vs_spherical_results}). Similarly, halos of the same shape, but different virial mass and concentration would also lead to different acceleration ratios. This NFW parameter-shape degeneracy (along with the fact that a 2D plot cannot fully describe the effects of four variables) is captured in figure \ref{triaxial_vs_spherical_results}. 

\begin{figure}
\centering
  \begin{subfigure}[b]{0.4855\textwidth}
  \centering
    \includegraphics[width=1.00\textwidth]{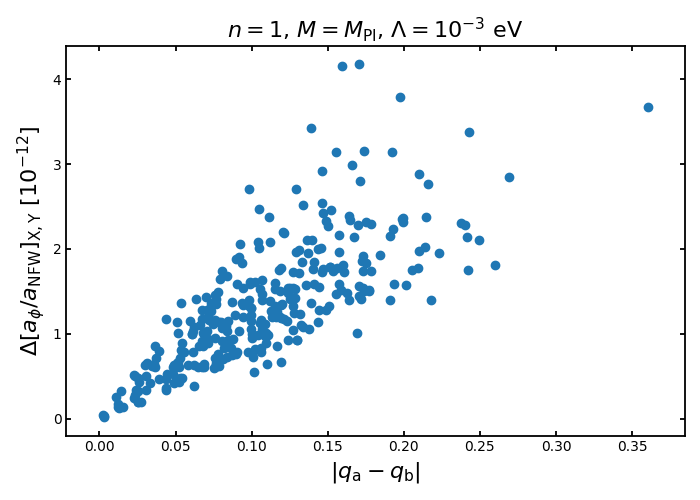}
    \label{axis_ratios_vs_acc_ratios:a1}
    
  \end{subfigure}
  \begin{subfigure}[b]{0.49\textwidth}
  \centering
    \includegraphics[width=1.00\textwidth]{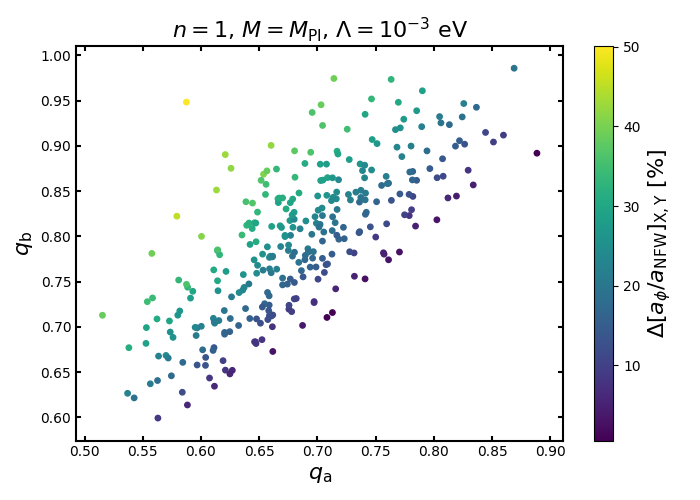}
    \label{axis_ratios_vs_acc_ratios:a2}
  \end{subfigure}
    \begin{subfigure}[b]{0.49\textwidth}
  \centering
    \includegraphics[width=1.00\textwidth]{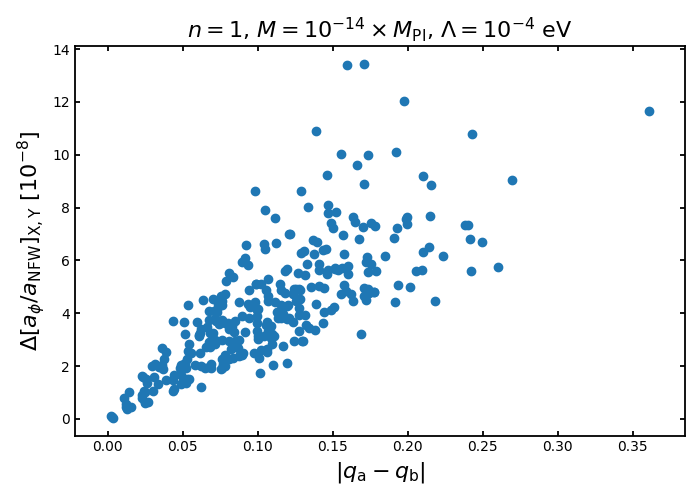}
    \label{axis_ratios_vs_acc_ratios:a3}
  \end{subfigure}
  \begin{subfigure}[b]{0.49\textwidth}
  \centering
    \includegraphics[width=1.00\textwidth]{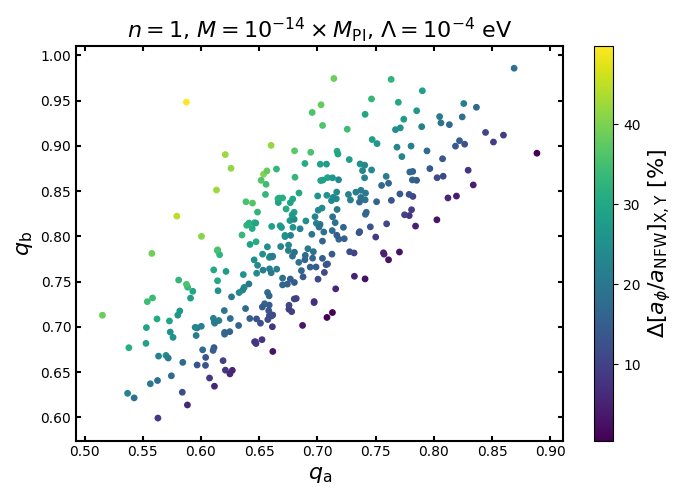}
    \label{axis_ratios_vs_acc_ratios:a4}
  \end{subfigure}
  \caption{The relationship between the axis ratios and the acceleration ratios measured at $r_{200}$ along the X and the Y axes. \textit{Left:} The absolute axis ratio difference vs. the corresponding X-Y axis acceleration ratio difference for different chameleon models. \textit{Right:} the axis ratios vs. the $a_{\phi}/a_{\rm NFW}$ ratio difference along the X and the Y axes for different chameleon models. The colorbar is based on the percentage difference between the ratios: $100 \% \times |a_{\rm ratio}^{X} - a_{\rm ratio}^{Y}|/a_{\rm ratio}^{X}$, with $a_{\rm ratio} \equiv a_{\phi}/a_{\rm NFW}$.}
  \label{axes_ratios_vs_acceleration}
\end{figure}

Another key feature of the triaxial NFW halos, as described in section \ref{nfw_halos}, is the expected X-Y axis acceleration ratio difference for equidistant measurements. As illustrated in figure \ref{triaxiality_effects}, we expect a significant difference between the chameleon (and NFW) acceleration, when measured at the virial radius. The relationship between the axis ratios $q_{a}$ and $q_{b}$ and the acceleration ratio is captured in figure \ref{axes_ratios_vs_acceleration}. The figure illustrates the key point that deviations from spherical symmetry lead to different values of the acceleration ratio when measured along the X and the Y axes. In the extreme case, the X-Y axis difference between the acceleration ratios can reach up to $50\%$. As before, there is a degeneracy between the shape parameters and the NFW properties.

Given the mentioned degeneracy between the NFW parameters and the axis ratios, we investigated the importance of each of these variables on the chameleon-to-NFW acceleration ratio. This was done by splitting the full cluster dataset into a \textit{training} (67\% of the full data) and \textit{test} (33\% of the full data) datasets. The training data was then used to train regression and decision tree based algorithms to correctly predict the acceleration ratio based on the four features: $M_{200}$, $c_{\rm NFW}$, $q_{a}$ and $q_{b}$. The prediction accuracy was tested on the test dataset. The algorithms were then used to estimate the feature importance (as described in section \ref{FEM_introduction}). The results are summarized in table \ref{feature_importance}. The key finding is that the changes to the virial mass have the most significant effect on the acceleration ratio (as expected), with the concentration parameter being the second most important variable. The shape parameters were found to be less important.

\begin{table}[]
\centering
\begin{tabular}{c|cccc}
\multicolumn{1}{c| }{} & Linear Regression  & Decision Trees & XGBoost  &  \\ \hline \hline
                    $M_{200}$      & 0.86  & 0.84 & 0.71  &  \\
                     $c_{\rm NFW}$ & 0.36   & 0.14 & 0.19 &  \\
                     $q_{a}$       & 0.12  & 0.01 & 0.06 & \\ 
                     $q_{b}$       & 0.03   & 0.01 & 0.04 & 
\end{tabular}
\caption{The results of the feature importance analysis. The values refer to the feature importance as evaluated by each algorithm. For linear regression this refers to the absolute value of the regression coefficients. For decision trees and XGBoost algorithms, the value refers to the feature importance based on the Gini importance (see section \ref{FEM_introduction}). The mean absolute error, $M_{E} = (\sum_{i}| y_{i} - y_{i}^{\rm pr}|)/n$, between the observed and the predicted values $y_{i}$ and $y_{i}^{\rm pr}$ for training (tr) and test (ts) data is then given by: $M_{E}^{\rm tr} = 0.27$, $M_{E}^{\rm ts} = 0.28$ (linear regression), $M_{E}^{\rm tr} = 0.13$, $M_{E}^{\rm ts} = 0.30$ (decision trees), $M_{E}^{\rm tr} = 0.04$, $M_{E}^{\rm ts} = 0.15$ (XGBoost). In all cases the test and the training datasets were normalized to have a mean value of 0 and a standard deviation of 1.  }
\label{feature_importance}
\end{table}

It should be noted, however, that the values listed in table \ref{feature_importance} should be interpreted with care as their meaning is algorithm-dependent. Also note that these values are only meaningful if the algorithms and the chosen features have enough predictive power. The corresponding mean absolute error values for the training and test datasets is given in the caption of table~\ref{feature_importance}.

\subsection{Triaxiality Results with the NFW Parameters Fixed}

\begin{figure}
\centering
  \begin{subfigure}[b]{0.49\textwidth}
  \centering
    \includegraphics[width=1.00\textwidth]{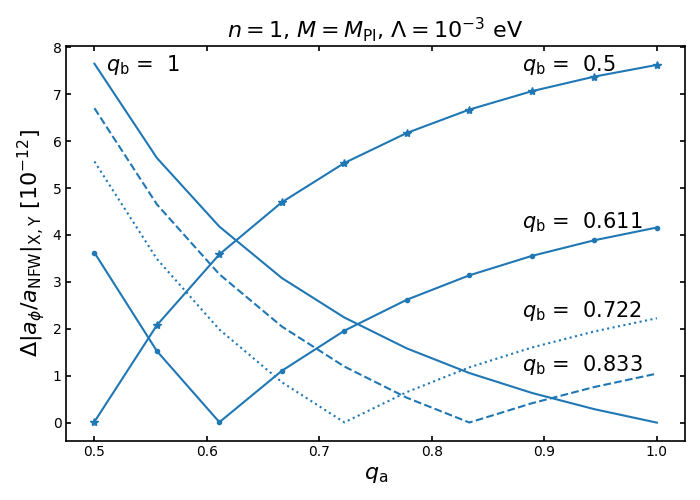}
    \label{triaxiality_NFW_fixed:a1}
    
  \end{subfigure}
  \begin{subfigure}[b]{0.49\textwidth}
  \centering
    \includegraphics[width=1.00\textwidth]{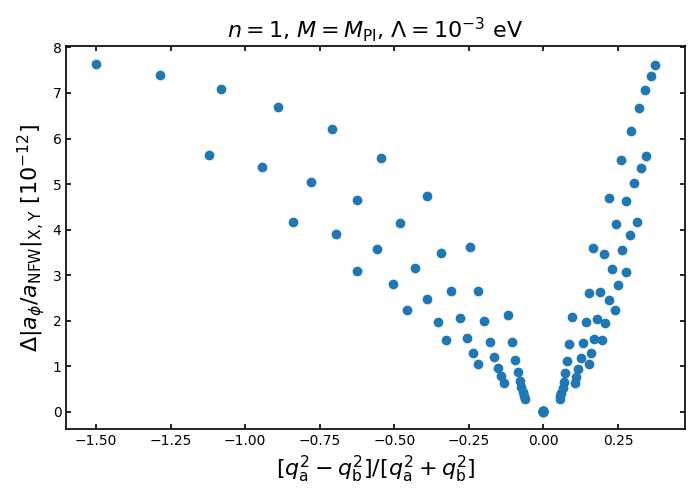}
    \label{triaxiality_NFW_fixed:a2}
  \end{subfigure}
\begin{subfigure}[b]{0.49\textwidth}
  \centering
    \includegraphics[width=1.00\textwidth]{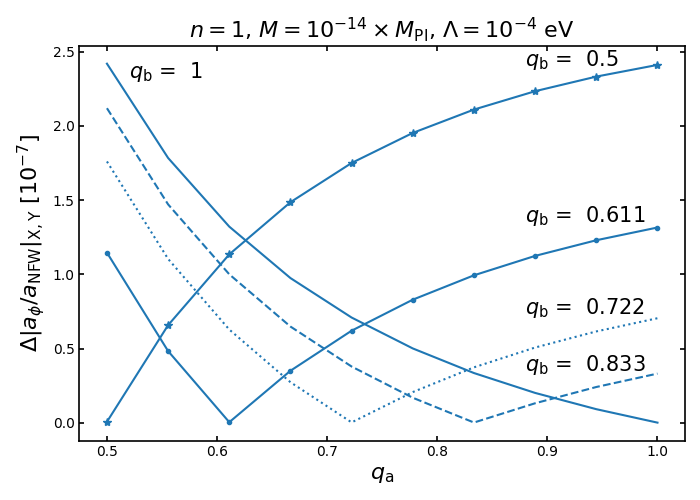}
    \label{triaxiality_NFW_fixed:a3}
    
  \end{subfigure}
    \begin{subfigure}[b]{0.49\textwidth}
  \centering
    \includegraphics[width=1.00\textwidth]{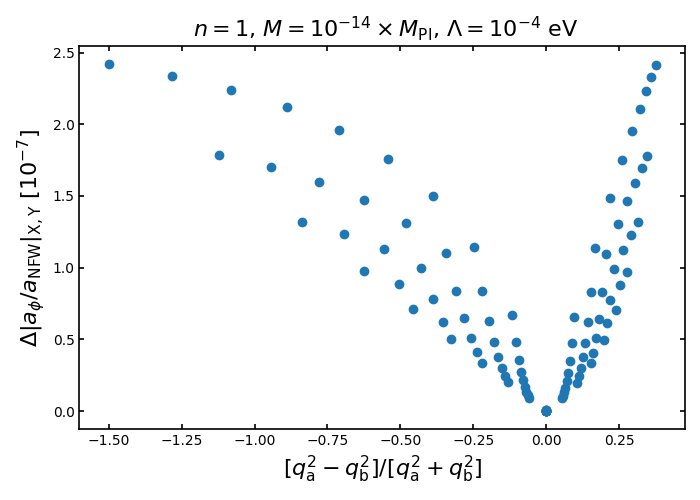}
    \label{triaxiality_NFW_fixed:a4}
  \end{subfigure}
  \caption{Triaxiality results for the NFW parameters fixed to the mean values in the dataset ($c_{\rm NFW} = 4.0$ and $M_{200} = 1.5 \times 10^{14} M_{\odot}$). \textit{Left:} the axis ratios vs. the equidistant chameleon-to-NFW acceleration ratios measured at the virial radius along the X and the Y axes for different chameleon models. \textit{Right:} eccentricity vs. the equidistant acceleration ratio for different chameleon models. }
  \label{triaxiality_results_NFW_fixed}
\end{figure}

As described in the previous subsection, interpreting the triaxial results is difficult due to the degeneracy between the NFW parameters and the axis ratios. In order to isolate the effects due to triaxiality alone, we repeated the previous analysis while keeping the NFW parameters fixed. Specifically, the virial mass and the concentration were set to the mean values found in our dataset: $c_{\rm NFW} = 4.0$ and $M_{200} = 1.5 \times 10^{14} M_{\odot}$. The calculations were done for the axis ratios in the realistic range of $q_{a},q_{b} \in [0.5,1.0]$. The investigation focused on how the chameleon-to-NFW acceleration ratio measured at equal distance (at $R = r_{200}$) along the X and the Y axes depends on the difference between the axis ratios. The results are summarized in figure \ref{triaxiality_results_NFW_fixed}, which illustrates the effect of the axis ratio difference on the chameleon-to-NFW acceleration ratios measured along the different axes. The intuition established in the previous subsection remains relevant here -- we find that higher deviations from spherical symmetry lead to a higher difference of the acceleration ratios. In other words, higher eccentricity leads to higher acceleration differences, as shown in the relevant subfigures. Figure \ref{triaxiality_results_NFW_fixed} also explores the results for lower values of $M$ and $\Lambda$, which, as we saw in the previous figures, highly increase the $a_{\phi}/a_{\rm NFW}$ values.  

\section{Conclusions}
\label{section:conclusions}

In this work we investigated how taking into account the density distribution of a galaxy cluster halo changes our estimate of chameleon screening and the strength of the chameleon fifth force. 
Specifically, we used the finite element method to solve the chameleon field equation for the T{\scriptsize HE} T{\scriptsize HREE} H{\scriptsize UNDRED} P{\scriptsize ROJECT} galaxy cluster density profiles. We explored different chameleon models with the main focus on models which are still allowed by the current observational constraints. For these models, the radial profiles of chameleon-to-NFW acceleration ratios were calculated. In addition, the effects due to NFW halo triaxiality were investigated. 

The key results indicate that the chameleon-to-NFW acceleration radial profile increases throughout the cluster monotonically. In other words, to maximise the chance of detecting the chameleon effects, one would have to focus on the outer regions of the galaxy cluster, i.e. at around the $r_{200}$ radius. It is important to note that this behaviour is controlled by the NFW radial acceleration profile, as the chameleon profile can behave differently depending on the $n$ parameter value. Specifically, $a_{\phi}(r)$ can decrease or increase towards the outer regions of the cluster depending on the value of $n$. Nonetheless, since the chameleon acceleration is significantly smaller than the corresponding NFW acceleration, the ratio of the two increases towards the virial radius in all studied cases.

A natural question is what our results imply for the observational searches for chameleon gravity. In this work we have shown that it is important to understand the internal structure and halo properties of galaxy clusters in order to calculate the screened fifth force correctly. Unfortunately, even with the halo properties taken into account, our results indicate that the fifth force in cluster sized halos would be very small and most likely not significant enough to be detected in future observational searches. More specifically, for all studied chameleon models we found $a_{\phi} \ll a_{\rm NFW}$. Even for the $n =1$, $M = 10^{-14} \times M_{\rm Pl}$ and $\Lambda = 10^{-4}$ eV model, which leads to the highest chameleon acceleration (amongst the models that are still allowed by the observational constraints), the chameleon-to-NFW acceleration ratio is around $10^{-8}-10^{-7}$. Note that the chameleon-to-NFW acceleration ratios are equal to the corresponding mass ratios, which means that in order to detect such small deviations in mass, extremely small measurement errors would be required. Some of the most accurate cluster mass data comes from the caustic mass measurements, which generally have relatively small errors when compared to other methods. As an example, in Ref.~\cite{Maughan2016} caustic and X-ray masses are compared for a selection of galaxy clusters. The findings show caustic mass errors ranging from sub-percent to $\sim 30\%$. However, it should be pointed out that in the same publication the authors find significant differences between the corresponding X-ray and caustic masses (generally less than $\sim 20\%$ difference). This shows that mass measurements are generally difficult. And hence an extremely small modified gravity signal would not be measurable and could easily be mistaken for various systematic measurement effects or possibly degenerate cluster astrophysical effects. The errors could be reduced by stacking galaxy clusters, however, this would average out the information about the cluster shapes. In conclusion, our results, when combined with the current observational constraints, show that the models that are still allowed by the data would not have any significant effect on galaxy cluster scales.

In addition, one might ask what are the implications for galaxy-sized halos. Since galaxies have significantly lower virial masses, one would expect that the acceleration ratios are higher in these systems. Running our codes for a typical galaxy halo with the virial radius of 100 kpc and masses of around $10^{10} \times M_{\odot}$ results in the acceleration ratios of around $10^{-5}$. For a more extreme example, one of the least massive galaxies that has been detected, Segue 2, has a mass of $\sim 10^{5} \times M_{\odot}$ with a half-light radius of 46 pc \cite{Kirby2013}. Running our codes for a halo of similar size (assuming a very small value of concentration of $\sim 2$) leads to acceleration ratio of around 0.2. This is significantly higher than the cluster scale results, however, it should be noted that galaxies such as Segue 2 are extremely faint and difficult to detect. In addition, the chosen concentration value might not be correct for such a galaxy. Nonetheless, the calculated value of the acceleration ratio is a good estimate of an upper limit given the currently available galaxy scale data. 

In previous work in Ref.~\cite{Burrage2018B} it was shown that deviations from spherical symmetry in sources of constant density can increase the acceleration of a test particle by an order one factor. In this work we expanded the research into deviations from spherical symmetry by focusing on halos of varying density. The key result is that triaxiality introduces directional effects on acceleration measurements that could be instrumental in observational searches for the modified gravity effects. In extreme cases, the Chameleon-to-NFW acceleration ratio could change by 50\% when measured along the X and the Y axes. More generally, the triaxiality of the NFW halos results in an angular dependence of the equidistant acceleration measurements. This dependence could be used to distinguish a modified gravity signal from degenerate astrophysical effects.   

Another important point regarding triaxiality effects is that the NFW parameters (concentration and the virial mass) derived by AHF used in this work are based on the assumption of spherical symmetry. However, it is known that real galaxy clusters are triaxial and hence, assuming a spherical NFW profile, while fitting triaxial halos leads to extra bias. As discussed in section \ref{full_triaxiality_results}, we do not expect the corresponding errors to be larger than around 10-20\%. Nonetheless, one natural direction for future work could be using a fully triaxial dataset (such that both the NFW parameters and the axis ratios are derived using a triaxial NFW profile), which would reduce the bias due to underlying spherical symmetry assumption and would lead to a more accurate description of triaxiality effects.  

It is also important to note that the techniques presented in this work are rather general and can be used to investigate other gravity models. SELCIE FEM solver, in particular, can be easily extended to solve the field equations of the symmetron and $f(R)$ models. In addition, similar approaches (i.e. the $\varphi$enics software package \cite{Braden2021}) have been used to explore Vainshtein screening. Similarly, the mentioned triaxiality effects would be relevant to any scalar field model, which depends on the variations of the density profile. 

There are a number of directions that we plan to explore in future work. The model that we used to describe clusters in this work (as spherical and triaxial NFW halos) is rather simplistic. For instance, undoubtedly the assumption of a continuous NFW density distribution breaks down at scales where effects due to individual galaxies become important. Ideally, we would like to investigate the halo substructure effects by employing more realistic density distributions. Similarly, the work described in this paper could easily be extended by exploring time varying density distributions. 

More generally, chameleon gravity is known to affect structure formation in multiple ways. As discussed in Ref.~\cite{Mitchell2019}, chameleon gravity is known to alter the mass-concentration relation. However, results obtained in this work indicate that (at least for the surviving chameleon models) the acceleration ratios are extremely small and hence we do not expect such effects to be significant in the context of structure formation. Nonetheless, even subtle effects could leave an imprint in the summary statistics of the large scale structure. Investigating such effects is beyond the scope of this work, but we do plan to study such effects by running chameleon dark matter-only and hydrodynamical simulations.

Another extension could be to study the full 3D effects of triaxiality. Interestingly, the results outlined in this work can be easily adapted to investigate chameleon effects at laboratory scales. Previous work (e.g. Ref.~\cite{Burrage2018B}) has explored the effects of screening in the context of vacuum chamber experiments. The techniques presented in this work could be applied at laboratory scales for sources with varying density. Lastly, another interesting avenue to investigate is the behaviour of the chameleon and other models in cosmic voids. Voids, being the largest under-densities in the Universe, offer a natural setting for testing screening mechanisms.

\appendix
\section{The Analytic Approximation}
\label{appendix_analytical_approx}

A key result outlined in this work is that inside a halo the chameleon field tracks the minimum of the effective potential for all the models explored in this paper. In other words, eq.~(\ref{phi_min}) can be used to approximate the field value with high accuracy. Note that this is only true for small values of the rescaling parameter: $\alpha \ll 1$. Or, in other words, the gradient of the field must be significantly smaller than the other terms in the field equation (eq.~(\ref{rescaled_field_equation})). 

\begin{figure}
\centering
  \begin{subfigure}[b]{0.49\textwidth}
  \centering
    \includegraphics[width=1.00\textwidth]{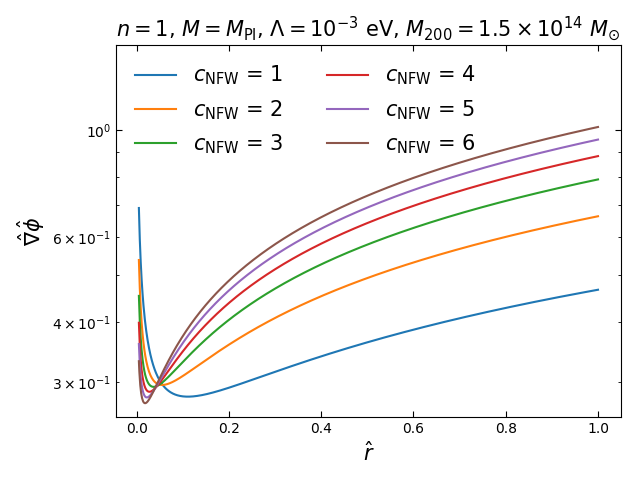}
    \label{grad_analytic:a1}
    
  \end{subfigure}
  \begin{subfigure}[b]{0.49\textwidth}
  \centering
    \includegraphics[width=1.00\textwidth]{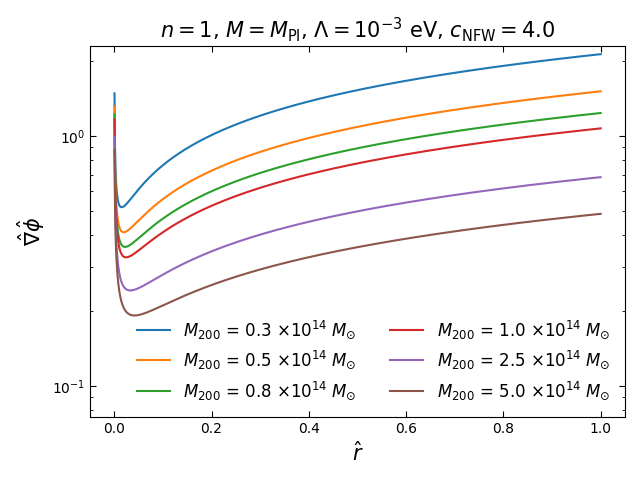}
    \label{grad_analytic:a2}
  \end{subfigure}
  \caption{The approximate analytic results for the dimensionless chameleon field gradient. \textit{Left:} the chameleon field gradient for different values of $c_{\rm NFW}$ with the virial mass set to the mean value from our dataset: $M_{200} = 1.5 \times 10^{14} M_{\odot}$. \textit{Right:} the chameleon gradient for different values of $M_{200}$ with $c_{\rm NFW} = 4.0$. The results were calculated using the domain size of $L = 3 \times r_{200}$.  }
  \label{gradient_analytic_results}
\end{figure}

Assuming that this approximation holds, one can derive the radial profiles for the field gradient for different NFW parameters. This is particularly simple in the dimensionless form (eq.~(\ref{rescaled_field_equation})), in which case the field is related to the density as described by equation \ref{phi_min_dimensionless}. The radial derivative of the field is then simply given by: 

\begin{equation}
    \hat{\nabla}\hat{\phi} = \frac{\partial \hat{\phi}}{\partial \hat{r}} = -\frac{\hat{\rho}'\hat{\rho}^{-\frac{(n+2)}{n+1}}}{n + 1},
    \label{analytic_gradient}
\end{equation}

\noindent with $\hat{\rho}'(\hat{r})$ as the radial derivative of the NFW profile rescaled by the background value. Figure \ref{gradient_analytic_results} shows the approximate analytic results for the chameleon gradient for different values of the NFW parameters. The key finding is that for very low values of $c_{\rm NFW}$, the chameleon gradient can be larger in the central regions than around the virial radius. Note that this is primarily due to the fact that $\lim_{\hat{r} \to 0} \hat{\rho}(\hat{r}) = \infty$. In real galaxy clusters, of course, we do not expect the density to grow infinitely in the central regions. However, it is possible that for very low concentration halos with high density in the central region, the chameleon gradient could be higher towards the centre than in the outer regions.

\section{Results Based on the Direct AHF Density Profile Fits}
\label{direct_fit_results}

\begin{figure}
\centering
\makebox[\linewidth][c]{%
  \begin{subfigure}[b]{0.49\textwidth}
  \centering
    \includegraphics[width=1.00\textwidth]{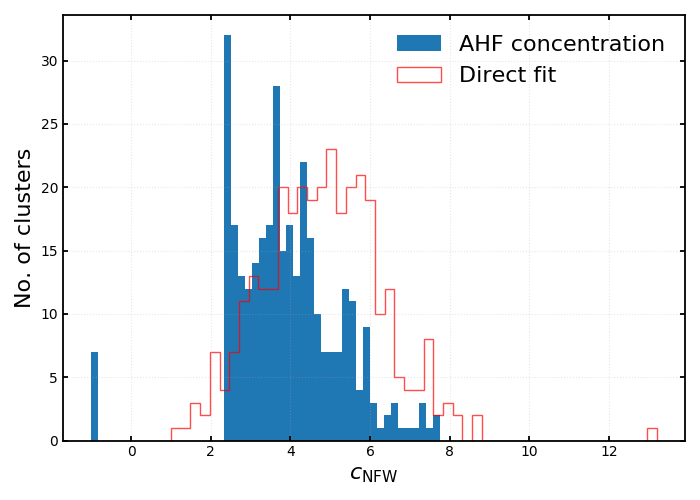}
    \label{concentration_comparison:a1}
    
  \end{subfigure}
  \begin{subfigure}[b]{0.49\textwidth}
  \centering
    \includegraphics[width=1.00\textwidth]{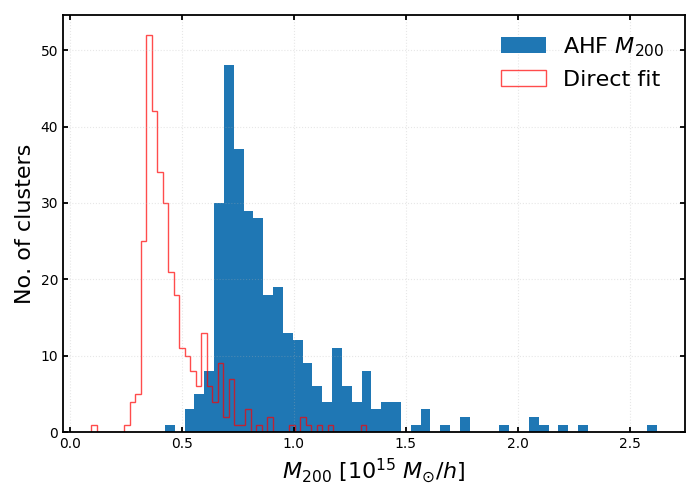}
    \label{concentration_comparison:a2}
  \end{subfigure}
  }
  \caption{The comparison of the AHF NFW parameters based on the maximum-to-virial velocity ratios (as described in Ref.~\cite{Prada2012}) and the direct fits of the radial density profiles. Note that the negative concentration values correspond to halos for which $c_{\rm NFW}$ could not be determined by the halo finder. These data points were not used in the analysis.}
  \label{concentration_comparison}
\end{figure}

\begin{figure}
\centering
  \begin{subfigure}[b]{0.49\textwidth}
  \centering
    \includegraphics[width=1.00\textwidth]{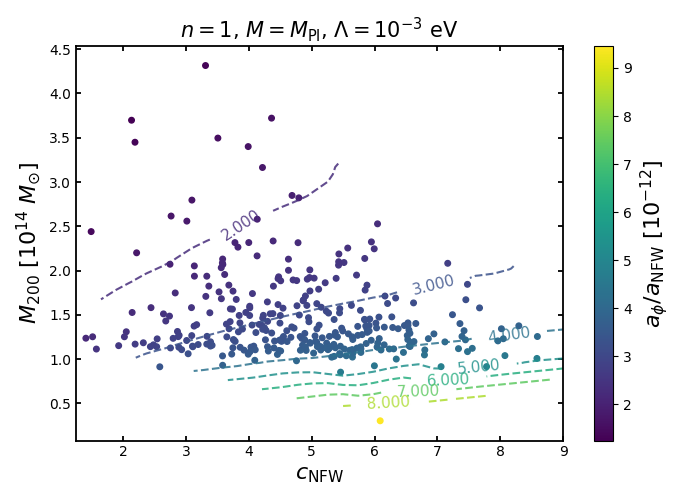}
    \label{direct_fit:a1}
    
  \end{subfigure}
  \begin{subfigure}[b]{0.49\textwidth}
  \centering
    \includegraphics[width=1.00\textwidth]{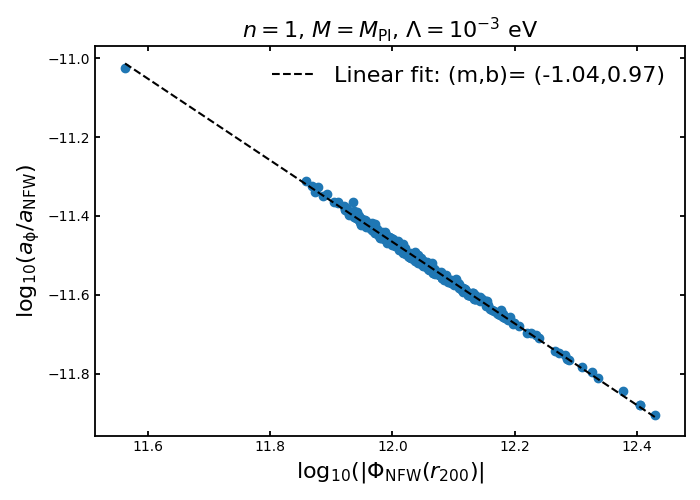}
    \label{direct_fit:a2}
  \end{subfigure}
\begin{subfigure}[b]{0.49\textwidth}
  \centering
    \includegraphics[width=1.00\textwidth]{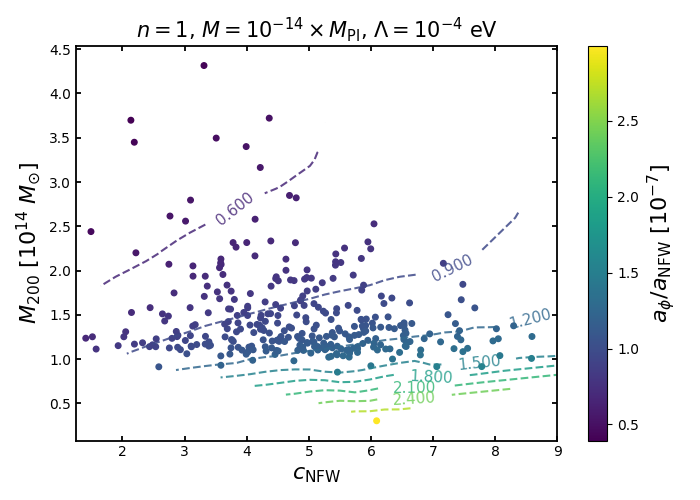}
    \label{direct_fit:a3}
    
  \end{subfigure}
    \begin{subfigure}[b]{0.49\textwidth}
  \centering
    \includegraphics[width=1.00\textwidth]{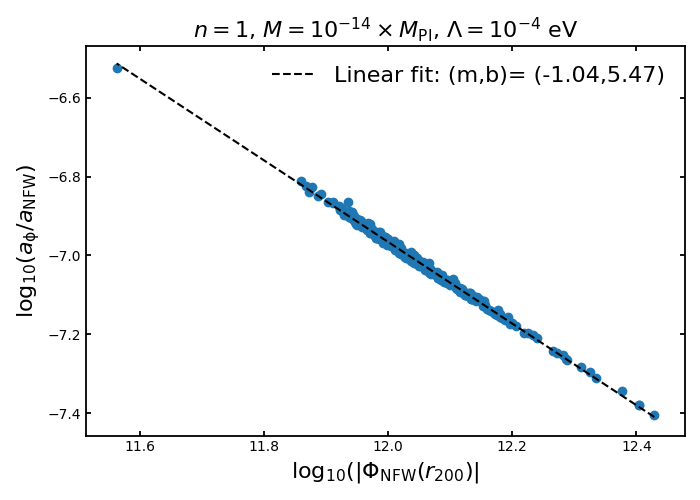}
    \label{direct_fit:a4}
  \end{subfigure}
  \caption{The chameleon-to-NFW acceleration ratio results using the direct concentration fit data. \textit{Top:} the chameleon-to-NFW acceleration ratio (at the virial radius) versus the NFW parameters and the NFW potential for the $n = 1$, $M= M_{\rm Pl}$, $\Lambda = 10^{-3}$ eV model. \textit{Bottom:} an analogous comparison for the $n = 1$, $M= 10^{-14} \times M_{\rm Pl}$, $\Lambda = 10^{-4}$ eV model. }
  \label{direct_fit_results_1}
\end{figure}

Here we present the results based on the concentration parameter values determined by directly fitting the AHF density profiles. As discussed in section \ref{The_dataset}, we expect the results to be generally similar to the results based on concentrations determined from the velocity ratios. The distribution of the NFW parameters based on the direct fits is shown in figure \ref{concentration_comparison}. Similarly, figure \ref{direct_fit_results_1} shows the comparison between the chameleon-to-NFW acceleration ratio and the NFW parameters along with the NFW potential. As expected, for both models shown in figure \ref{direct_fit_results_1}, the results are very similar to the analogous results based on the velocity ratio concentration values. The only difference of note is that the direct fit concentration values are slightly larger than the analogous values based on the velocity ratios. Similarly, the direct fit virial mass values are slightly lower than the analogous values based on the velocity ratios. Otherwise, the results are nearly identical.

\acknowledgments

The calculations done in this work would not be possible without the generous support by the Sciama HPC cluster at the University of Portsmouth. 

This work has also been made possible by T{\scriptsize HE} T{\scriptsize HREE} H{\scriptsize UNDRED} P{\scriptsize ROJECT} collaboration (https://the300-project.org). The simulations used in this paper have been performed in the MareNostrum Supercomputer at the Barcelona Supercomputing Center, thanks to CPU time granted by the Red Espa\~{n}ola de Supercomputaci\'{o}n. As part of T{\scriptsize HE} T{\scriptsize HREE} H{\scriptsize UNDRED} P{\scriptsize ROJECT} project, this work has received financial support from the European Union’s Horizon 2020 Research and Innovation programme under the Marie Sklodowskaw-Curie grant agreement number 734374, the LACEGAL project. 

We would also like to sincerely thank Robert Mostoghiu, Frazer Pearce and Kathy Romer for the enlightening discussions on galaxy clusters. In addition, we thank Paul Giles and David Turner for their help with interpreting and understanding the newest galaxy cluster observational constraints.  

Clare Burrage and Andrius Tamosiunas are supported by a Research Leadership Award from The Leverhulme Trust. Chad Briddon is supported by the University of Nottingham. Weiguang Cui is supported by the European Research Council under grant number 670193 and by the STFC AGP Grant ST/V000594/1. He further acknowledges the science research grants from the China Manned Space Project with NO. CMS-CSST-2021-A01 and CMS-CSST-2021-B01. Adam Moss is supported by a Royal Society University Research Fellowship.

\bibliography{journal_names,refs} 
\bibliographystyle{JHEP}

\end{document}